\documentclass{article}

\usepackage{PRIMEarxiv}
\usepackage{amsmath}
\usepackage{subfig}
\usepackage[utf8]{inputenc} 
\usepackage[T1]{fontenc}    
\usepackage{hyperref}       
\usepackage{url}             
\usepackage{booktabs}       
\usepackage{amsfonts}        
\usepackage{nicefrac}       
\usepackage{microtype}      
\usepackage{lipsum}
\usepackage{upgreek}
\usepackage{fancyhdr}       
\usepackage{graphicx}       
\usepackage{multirow}
\graphicspath{{media/}}     

\title{Performance Analysis of a Prime-Parameterized Fibonacci Spiral-Based Optical Phased Array
}

\author{Anantha Kedar Sarma Inampudi\textsuperscript{1,2},
  Anjali A R\textsuperscript{1},
  Pranabendu Ganguly\textsuperscript{1},
  Syamsundar De\textsuperscript{1}\thanks{Corresponding author: \href{mailto:syamsundarde@atdc.iitkgp.ac.in}{syamsundarde@atdc.iitkgp.ac.in}} \\
  \\
  \textsuperscript{1}Advanced Technology Development Centre, Indian Institute of Technology Kharagpur, India \\
  \textsuperscript{2}Department of Physics, Indian Institute of Technology Kharagpur, India 
}

\begin{document}
\maketitle

\begin{abstract}
Optical phased arrays (OPAs) are a promising technology for realizing fast and on-chip non-mechanical beam steering. In this work, we propose and analyze the performance of a non-uniformly spaced antenna arrangement based on the Fibonacci Spiral. A unique prime-number-based parameterization for antenna positioning and a tunable positional-control parameter ($\alpha$) are introduced. We show that, depending on the intended application of the OPA, by adjusting $\alpha$, we can achieve $\approx$ 56,562 resolvable points with 93 antennas arranged according to the prime-parametrization. To the best of our knowledge, this result exceeds the reported values in existing literature for comparable non-redundant array configurations. We analyze the robustness of this design by evaluating the sensitivities of the three key performance metrics of an OPA: side-lobe suppression ratio (SLSR), field of view (FoV), and the far-field full width at half-maximum (FWHM), to the random disturbances in antenna positions that may occur for practical implementation on a photonic chip.
\end{abstract}


\section{Introduction}
\begin{figure}
    \centering
    \includegraphics[width=0.7\linewidth]{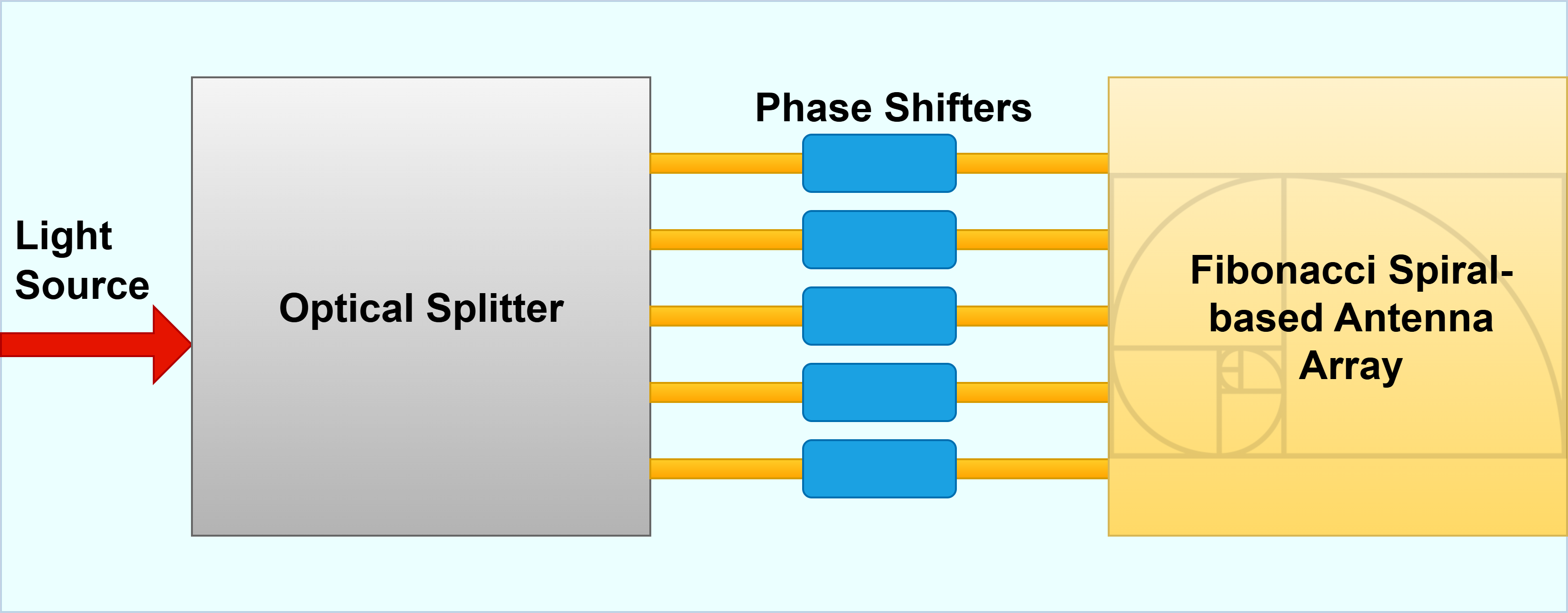}
    \caption{Schematic of the proposed Fibonacci-prime spiral-based optical phased array (OPA), illustrating five phase shifters for simplicity. The background Fibonacci spiral(yellow region) represents the bio-inspired geometric foundation of the antenna arrangement.}
    \label{fig:1}
\end{figure}

The use of an array of antennas for digital communication experiments dates back to the early 1900s, and since then, phased-array architectures have been widely used in radar systems\cite{N1}. The extension of this concept into optics, leading to optical phased arrays (OPAs), began in the second half of the 20th century \cite{N2}. OPAs work on the principle of coherent interference of the light emitted from the antennas to form a beam in the far-field, and the phase differences among them help in steering the beam \cite{N2}. Conventionally, OPAs rely on a periodic arrangement of antennas \cite{N3,N4}. However, the number of resolvable points ($N_{o}$), defined as the ratio of field of view of the OPA and the full width at half-maximum of its far-field beam, in a periodic OPA scales linearly with the number of antennas ($N$) \cite{N19}. This limitation constrains both scalability and compactness of OPAs. Arranging the antennas with non-uniform spacing can address scalability issues and enable beam steering to a larger set of distinct spatial points using fewer antennas. Various types of non-uniform arrangements have previously been explored\cite{N5,N6,N7,N8,N9}. Optimization algorithms, such as particle swarm optimization, have also been explored recently for designing aperiodic OPAs\cite{N16}. Non-redundancy in the arrangement of antennas plays an essential role in obtaining a narrower beam at the far-field plane, as reported by T. Fukui \textit{et al.}\cite{N7} using a mathematically non-redundant configuration in the form of a Costas Array. In this work, we present a nature-inspired alternative that exploits the inherent non-redundancy of the Fibonacci spiral. We further improve the Fibonacci design by including prime-number-based parameterization, and also provide a tunable angular antenna positioning parameter ($\alpha$) that enables the design to be flexible, allowing for the optimization of certain OPA parameters according to application requirements. This configuration achieves a degree of spatial non-redundancy comparable to that of, for example, the Costas array, while enabling a compact layout and tolerance to fabrication imperfections. Although the Fibonacci spiral has been studied in ultrasound phased arrays\cite{N17}, its application in OPAs remains unexplored. In addition, the far-field scattering properties of prime number arrays of metal nanoparticles to achieve broadband enhancement and localization of plasmonic fields were studied by Forestiere \textit{et al}.\cite{N18}. However, to the best of our knowledge, there are no prior reports that utilize a prime number parameterization for OPAs. In particular, the present approach combines prime number parameterization with a Fibonacci-spiral antenna arrangement and a tunable antenna positioning via $\alpha$. It hence offers a tunable trade-off among the side-lobe suppression ratio (SLSR), the field of view (FoV), the full width at half-maximum  (FWHM), and the number of resolvable points ($N_o$) in the far field. \\
The physical realization of OPAs can be achieved through various methods like liquid crystals, metasurfaces, and photonic integrated circuits (PICs). Nonetheless, PIC-based OPAs have been extensively explored due to their CMOS compatibility, compactness, and fast response times \cite{N10}. The schematic of our proposed OPA, which can potentially be implemented on a PIC platform, is shown in Fig.\,\ref{fig:1}. This block diagram shows only five phase shifters for simplicity. Light from a coherent source is divided using optical splitters, typically multimode interferometers (MMIs) or Y-splitter networks, and passed through individually tunable phase shifters to control the beam direction. The outputs are then routed through waveguides into the antenna array, here, the Fibonacci spiral-based array of antennas, such as grating antennas. 

In this work, we simulated the Fibonacci Spiral-based OPA employing various numbers of antennas arranged according to the prime-parameterization, and analyzed its far-field performance characteristics, using Python. In addition, we studied two different antenna configurations, obtained by varying the tunable control parameter $\alpha$, depending on the intended application of the OPA. For applications that require a high SLSR but can compromise on the number of resolvable points, Configuration 1 corresponding to $\alpha=0.45$ is suitable. On the other hand, for applications that prioritize a larger number of resolvable points, Configuration 2 corresponding to $\alpha=0.75$ can be considered. As an example, we have shown that for 93 antennas, Configuration 1 yields an SLSR of 9.64 dB and 13.78 dB in the two far-field directions $\theta,\psi$, respectively, and $\approx$14,086 resolvable points. In contrast, Configuration 2 results in $\approx$56,562 resolvable points but with a compromised SLSR of 12.09 dB and 5.18 dB in the $\theta,\psi$ directions. Nevertheless, in either of the configurations, we show that the number of resolvable points scales well above the quadratic dependence on the number of antennas ($N_{0}>N^{2}$) as expected in non-redundant OPAs. To incorporate fabrication imperfections for the practical realization of OPAs, we intentionally offset the antenna positions with random noise and investigate its impact on the SLSR, FoV, and FWHM. We find that the introduction of positional deviations has maximal impact on the SLSR along the $\psi$-axis, e.g., approximately 4\% positional variation leads to a standard deviation of around 2 dB in the $\psi$-axis SLSR.

\section{Principle and Design}
\label{sec:2}
Let E(\textbf{r}) and F($\boldsymbol{\xi}$) represent the electric field distributions at the near-field (NF) and far-field (FF) planes, respectively. Here, the vector \textbf{r} = (x,y) represents the coordinates at the NF plane and $\boldsymbol{\xi}$ = ($\theta,\psi$) denotes the lateral and longitudinal angles at the far-field plane. Let u(\textbf{r}) represent the electric field distribution from a single optical antenna in the vicinity of the emission plane. We follow the mathematical framework developed in Ref.\cite{N7}. For completeness, we reproduce the key expressions here:
\begin{equation}
    E(\textbf{r}) ~ = ~ \sum_{n=1}^N C_n u(\textbf{r}-\textbf{r}_n)
    \label{eq:1}
\end{equation}
Here, $C_n = |C_n|e^{i\phi_n}$ is the complex amplitude of the light emitted from $n$th optical antenna. Since the FF pattern is the Fourier transform of the NF pattern, we obtain:
\begin{equation}
    F(\boldsymbol{\xi}) ~ = ~ \iint E(\textbf{r})exp[ik_0(\textbf{r}.\boldsymbol{\xi})]d^2\textbf{r}
    \label{eq:2}
\end{equation}
Here, $k_0 = \frac{2\pi}{\lambda}$ is the wavenumber of light. In this work, we have considered $\lambda ~=~ 1550\,\textrm{nm}$.
\\ By substituting Equation \ref{eq:1} into Equation \ref{eq:2}, we obtain the following:
\begin{equation}
     F(\boldsymbol{\xi}) ~ = ~ U(\boldsymbol{\xi}) \sum_{n=1}^NC_nexp[ik_0(\textbf{r}_n.\boldsymbol{\xi})]
     \label{eq:3}
\end{equation}
Where, $U(\boldsymbol{\xi}) ~ = ~ \iint u(\textbf{r})exp[ik_0(\textbf{r}.\boldsymbol{\xi})]d^2\textbf{r}$ denotes the Fourier transform of $u(\textbf{r})$

\begin{equation}
    \implies F(\theta,\psi) ~ = ~ U(\theta,\psi)\sum_{n=1}^N C_nexp[ik_0(x_n\theta+y_n\psi)]
    \label{eq:4}
\end{equation}
From this, we obtain the far-field intensity as:
\begin{equation}
    I(\theta,\psi) ~ = ~ |F(\theta,\psi)|^2 ~ = ~ |U(\theta,\psi)|^2\sum_{n=1}^N\sum_{m=1}^N C_nC_m^*exp[(\textbf{r}_n-\textbf{r}_m).\boldsymbol{\xi}]
    \label{eq:5}
\end{equation}
 In this work, we have considered the electric field distribution from a single antenna at the NF plane, u(\textbf{r}), as a Gaussian with a spot size of $w ~ = ~ 2 ~ \mu m$. Therefore,
\begin{equation}
    u(\textbf{r}) ~=~ \frac{1}{2\pi w^2} exp\Big(\frac{-r^2}{2w^2}\Big)
    \label{eq:6}
\end{equation}

\begin{equation}
    \implies U(\theta,\psi) ~=~ exp\Big[-\frac{1}{2}(k_0w)^2(\theta^2+\psi^2)\Big] 
    \label{eq:7}
\end{equation}
The far-field intensity is calculated by substituting Equation \ref{eq:7} into Equation \ref{eq:5}. A more detailed derivation is given in Ref. \cite{N7}. We adopt this framework to analyze the performance of our Fibonacci-spiral-based antenna array as detailed below.

\subsection{Fibonacci design}
The Fibonacci spiral equation is given by:
\begin{equation}
    r ~=~ a_p \phi^{\frac{2\theta_p}{\pi}} 
    \label{eq:8}
\end{equation}
Here,
\begin{itemize}
    \item $(r,\theta_p)$ denote the polar coordinates of the position of an antenna
    \item $a_p$ is a scaling constant 
    \item $\phi ~=~ \frac{1+\sqrt5}{2}$ is the golden ratio
\end{itemize}
$a_p$ and $\theta_p$ are the two parameters that determine the position of an antenna in the Fibonacci spiral. For a particular $a_p$, the antennas are arranged along the arc by varying $\theta_p$. 
\\We varied the radius scaling parameter $a_p$ using the natural numbers set: $a_p ~\in~ \{1,2,3,4...\} $.
\\For the angular parameter $\theta_p$, we have heuristically developed a prime number parameterization given as:
\begin{align}
    \theta_p =
    \begin{cases}
    0,\pi ~~~~~~~~~~ &\text{for}~a_p~=~1\\
    \frac{j\pi}{k}&\text{for}~a_p~>~1 
    \end{cases}
    \label{eq:9}
\end{align}
\begin{align*}
    \text{where,}~k~\in~\{2,3,5,7,...\}~\text{denotes the prime numbers set, and }j~\in~\{1,2,3,...,k-1\} 
\end{align*}
We restricted $k$ to prime numbers to improve the non-redundancy in the arrangement. Using these constraints in $a_p$ and $\theta_p$, we have populated the array of antennas. To illustrate the procedure, the polar coordinates for 6 of the antennas were taken from:
\begin{align*}
    a_p ~=~ 5, ~\theta_p ~\in~ \{\frac{\pi}{7},\frac{2\pi}{7},\frac{3\pi}{7},\frac{4\pi}{7},\frac{5\pi}{7},\frac{6\pi}{7}\}
\end{align*}
The Cartesian coordinates of each antenna are obtained from $(x,y) ~=~ (rcos\theta_p, rsin\theta_p)$. The coordinates were scaled with $15\,\mu m$ to provide a realistic design.\\
To further optimize the design, we have introduced an additional angular position variation $\Delta\theta_p$ into the Equation \ref{eq:8}. From this, the modified spiral equation can be given as:
\begin{equation}
    r ~=~ a_p\phi^{\frac{2\theta_p\Big(1+\frac{\Delta\theta_p}{\theta_p}\Big)}{\pi}}
    \label{eq:10}
\end{equation}
Changing this angular position control parameter, $\alpha ~=~\frac{\Delta\theta_p}{\theta_p}$, allows us to obtain different OPA configurations adapted to the optimal values of any particular far-field parameters specific to application requirements. In particular, we have considered two  $\alpha$-dependent configurations: 1) for maximum SLSR, and 2) for maximum resolvable points.

\begin{figure}[htbp]
  \centering
  \subfloat[\label{fig:2a}]{
    \includegraphics[width=0.48\textwidth]{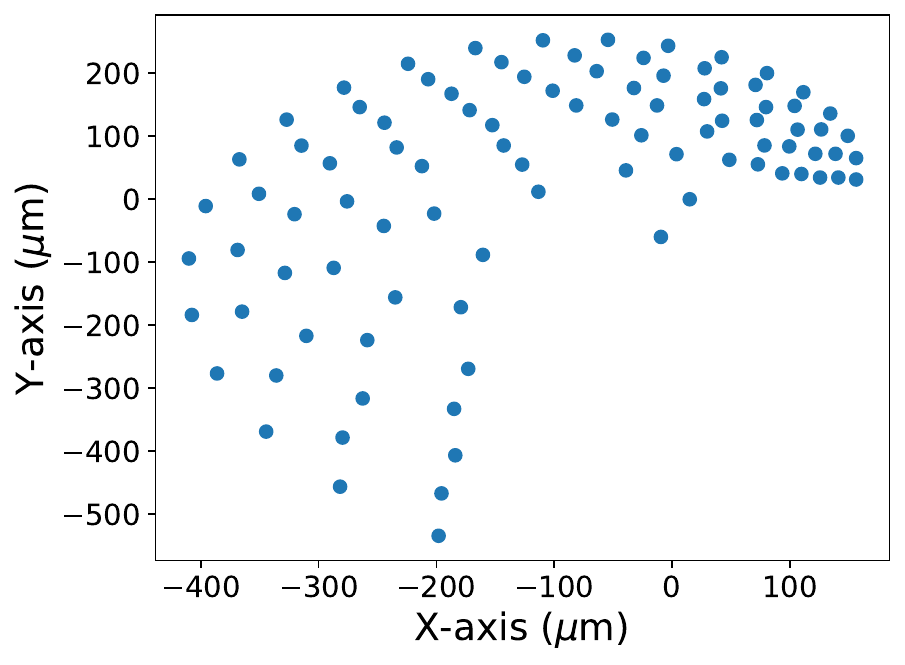}
  }
  \hfill
  \subfloat[\label{fig:2b}]{
    \includegraphics[width=0.48\textwidth]{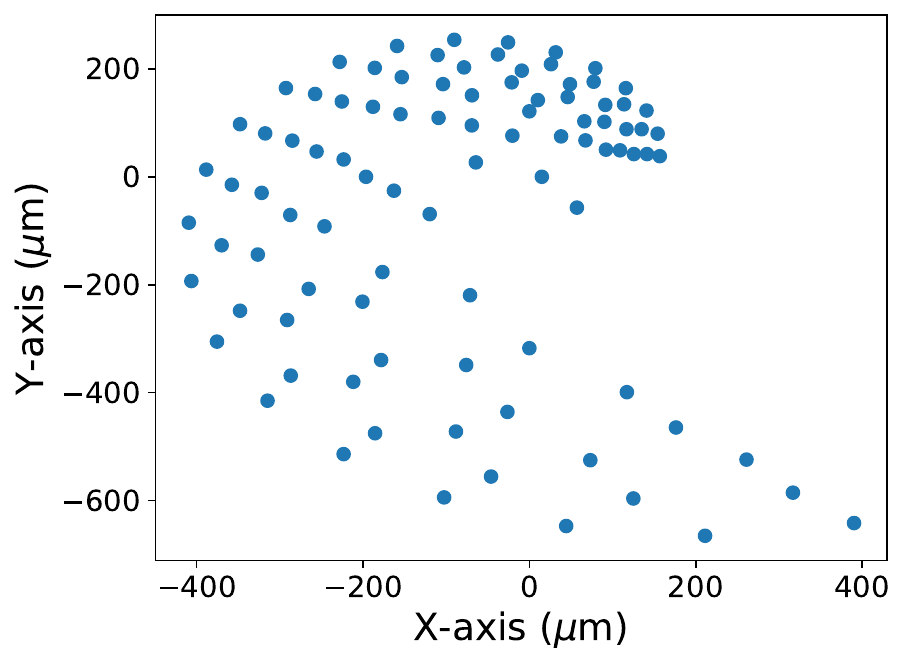}
  }
  
  \vskip\baselineskip  
  \subfloat[\label{fig:2c}]{
    \includegraphics[width=0.48\textwidth]{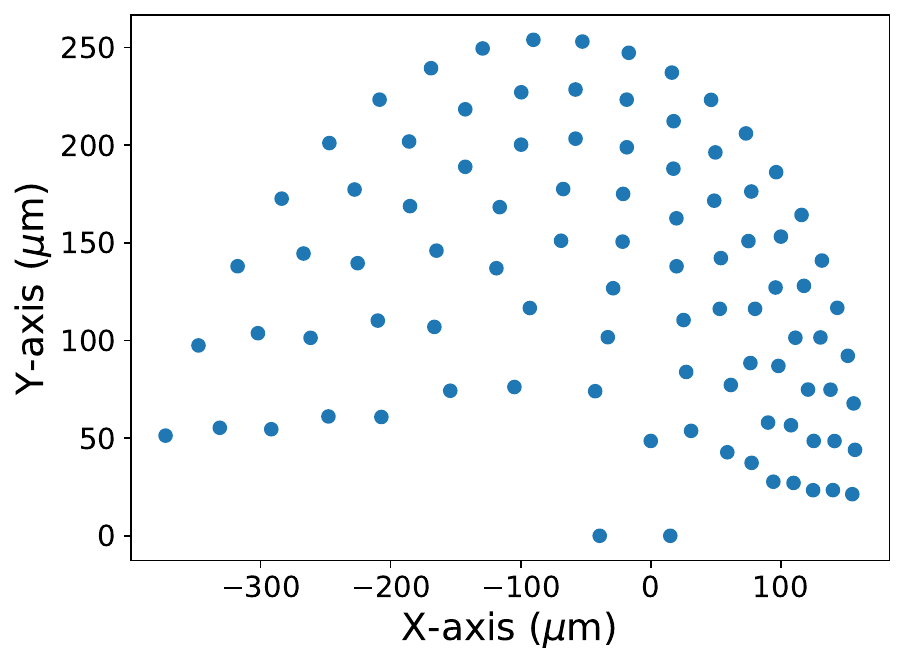}
  }
  \caption{Arrangement for 93 antennas in (a) configuration 1 ($\alpha ~=~ 0.45$) (b) configuration 2 ($\alpha ~=~ 0.75$) (c) a configuration with $\alpha ~=~ 0$. The coordinate axes represent antenna positions in the x-y plane.}
  \label{fig:2}
\end{figure}

\section{Simulation Results}
Using the theory developed in Section \ref{sec:2}, we have analyzed various far-field parameters for the Fibonacci-prime arrangement in both configurations mentioned above.  Specifically, Configuration 1, corresponding to the maximum SLSR, is obtained with $\alpha ~=~ 0.45$, and Configuration 2, corresponding to the highest number of resolvable points, is obtained with $\alpha ~=~0.75$. These two configurations for 93 antennas with a minimum spacing of $15~\mu m$ are shown in Figures \ref{fig:2a} and \ref{fig:2b}, respectively. For comparison, we show the arrangement for 93 antennas with $\alpha ~=~ 0$ in Fig.\,\ref{fig:2c}.  Here, we have executed the performance analysis of the proposed OPA design, for example, for 93 antennas, since this corresponds to the arrangement of antennas that span 10 complete arcs, beginning with $a_p ~=~1$ for the first arc and ending with $a_p ~=~10$ for the last arc. 

\begin{figure}[htbp]
    \centering
  \subfloat[\label{fig:3a}]{
    \includegraphics[width=0.48\textwidth]{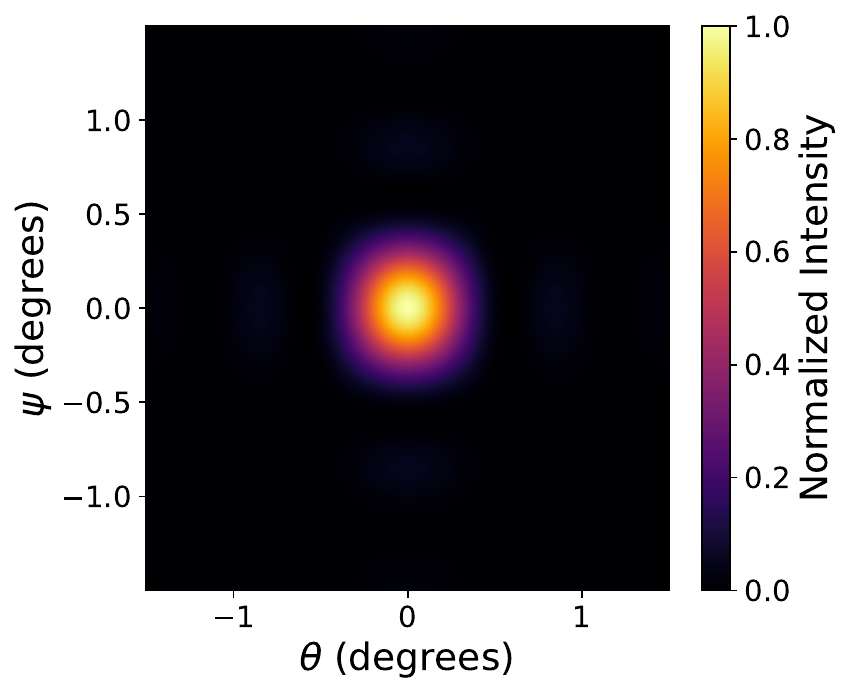}
  }
  \hfill
  \subfloat[\label{fig:3b}]{
    \includegraphics[width=0.48\textwidth]{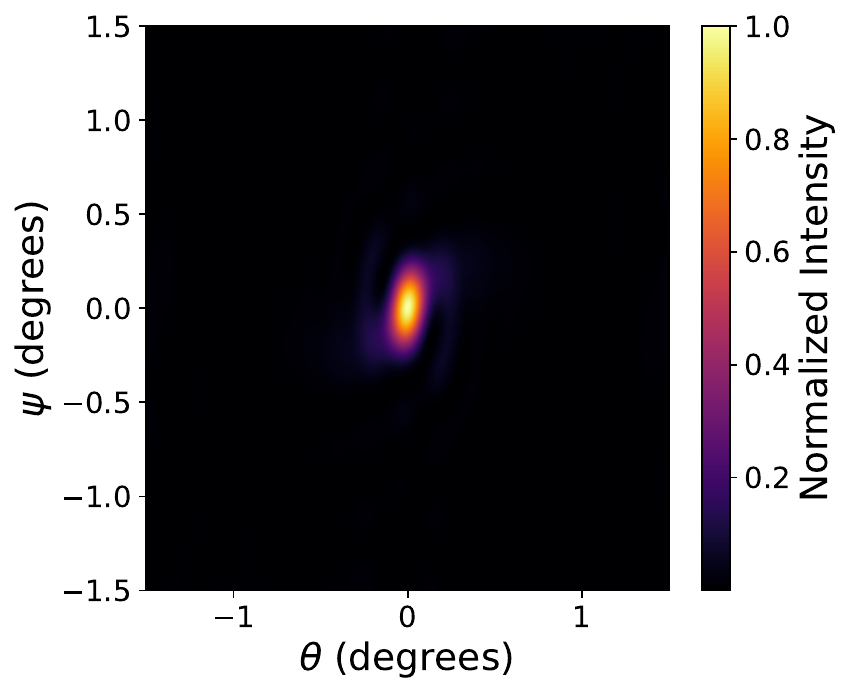}
  }
    \caption{Simulated far-field radiation pattern for (a) uniform arrangement of 100 antennas. (b) an arrangement of 93 antennas in Fibonacci-prime layout with $\alpha ~=~ 0$.}
    \label{fig:3}
\end{figure}

To start with, in Fig.\,\ref{fig:3a}, we show the far-field pattern for a uniform arrangement of 100 antennas with a $15~\mu m$ spacing between neighbouring antennas, and compared with the Fibonacci-prime arrangement of 93 antennas with $\alpha ~=~ 0$ given in Fig.\,\ref{fig:3b}. The far-field pattern of the uniform arrangement corresponds to an FWHM of $0.4498^\circ \times 0.4498^\circ$, while the pattern is narrowed more than three times with an FWHM of $0.1499^\circ \times 0.2998^\circ$ for the Fibonacci-prime arrangement. This validates the proposed design, confirming that the Fibonacci-prime aperiodic layout enables a significantly tighter beam confinement than a periodic design. This confinement can be attributed to the enhanced spatial non-redundancy of the Fibonacci-prime layout, which suppresses repeated interferences that typically broaden the main lobe in a periodic array.

\begin{figure}[htbp]
  \centering
  \subfloat[\label{fig:4a}]{
    \includegraphics[width=0.48\textwidth]{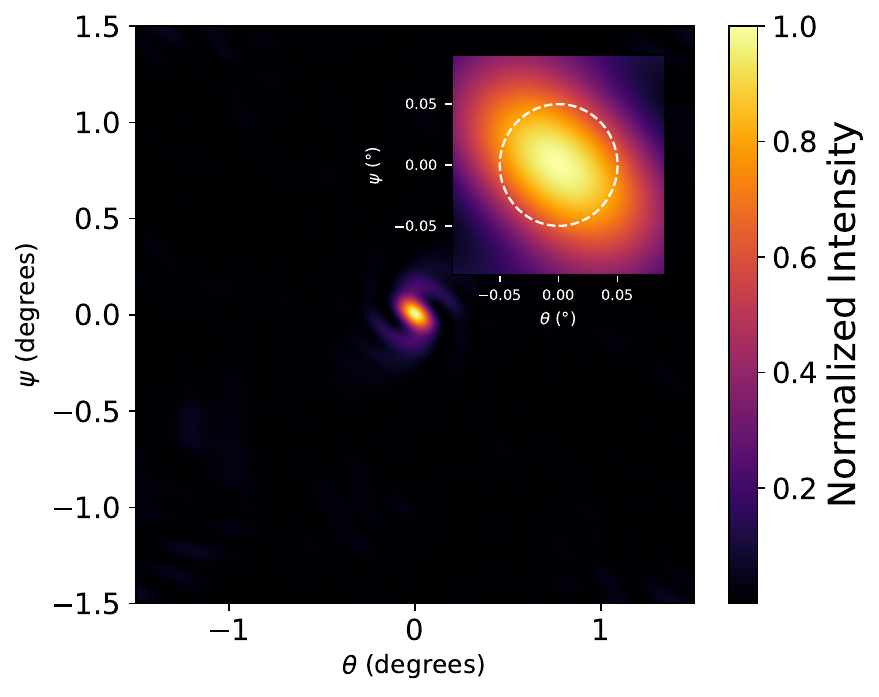}
  }
  \hfill
  \subfloat[\label{fig:4b}]{
    \includegraphics[width=0.48\textwidth]{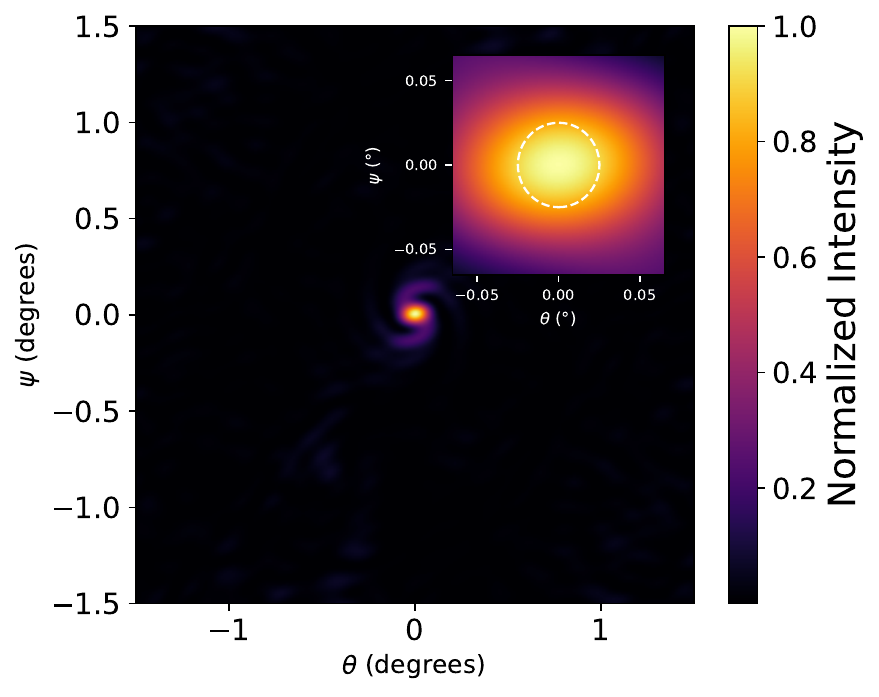}
  }
  \caption{Simulated two-dimensional far-field pattern for the Fibonacci-prime array with 93 antennas: (a) configuration 1 ($\alpha ~=~ 0.45$) (b) configuration 2 ($\alpha ~=~ 0.75$). Color scale indicates normalized intensity in the $\theta-\psi$ plane. Insets display magnified views around the main lobe region to highlight beam confinement. White dashed contour in the insets mark the regions of FWHM}
  \label{fig:4}
\end{figure}

Subsequently, we simulated the far-field patterns for both Configuration 1 ($\alpha ~=~ 0.45$) and 2 ($\alpha ~=~ 0.75$) using 93 antennas and compared their far-field intensities as shown in Figures \ref{fig:4a} and \ref{fig:4b}, respectively.

Configuration 1 resulted in a beam with an FWHM of $0.0999^\circ \times 0.0999^\circ$, while the FWHM of Configuration 2 is $0.050^\circ \times 0.050^\circ$. Both these configurations of the Fibonacci-prime arrangement represent a significant narrowing in FWHM compared to that of $\alpha ~=~ 0$ as in Fig.\,\ref{fig:3b}, indicating that the $\alpha$ parameter serves as an effective control for an optimal antenna positioning to get a tighter far-field beam confinement.

\begin{figure}[htbp]
  \centering
  \subfloat[\label{fig:5a}]{
    \includegraphics[width=0.7\textwidth]{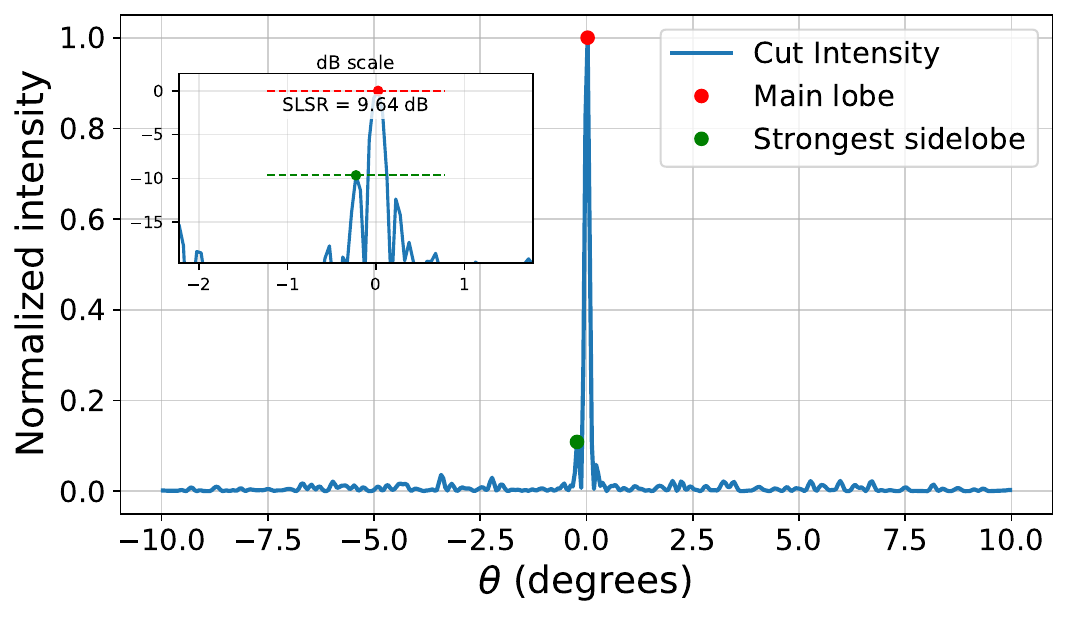}
  }
  \hfill
  \subfloat[\label{fig:5b}]{
    \includegraphics[width=0.7\textwidth]{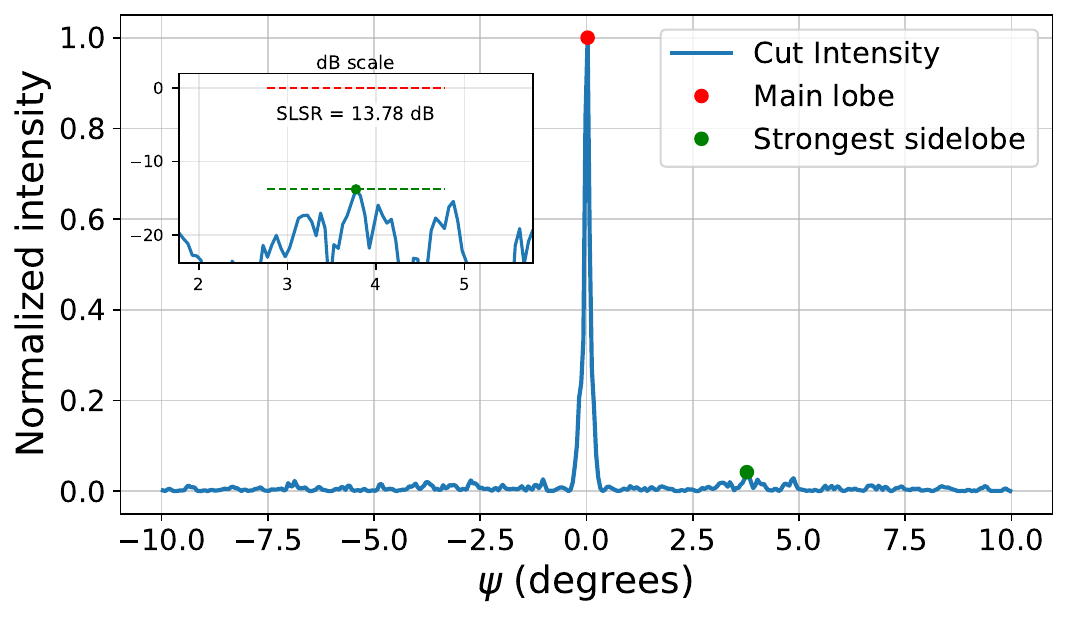}
  }
  \caption{Simulated one-dimensional far-field patterns for the Fibonacci-prime array with 93 antennas in configuration 1: (a) $\theta$-direction (b) $\psi$-direction. Red and green markers indicate the main lobe and the strongest side lobe, respectively. Insets present zoomed views around the strongest side lobe in a logarithmic (dB) scale}
  \label{fig:5}
\end{figure}

\begin{figure}[htbp]
  \centering
  \subfloat[\label{fig:6a}]{
    \includegraphics[width=0.7\textwidth]{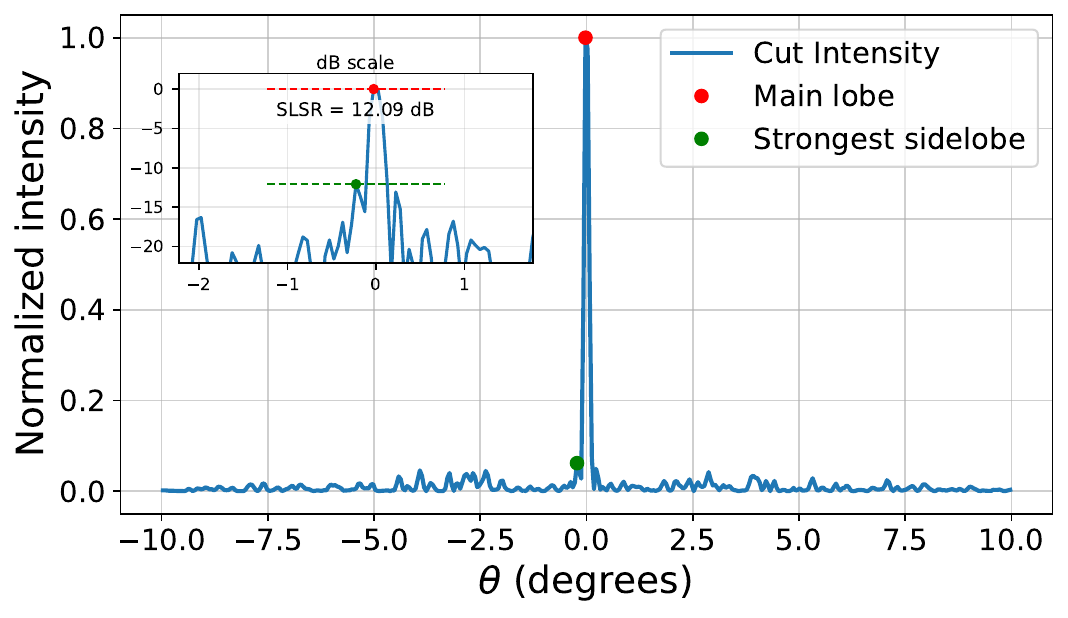}
  }
  \hfill
  \subfloat[\label{fig:6b}]{
    \includegraphics[width=0.7\textwidth]{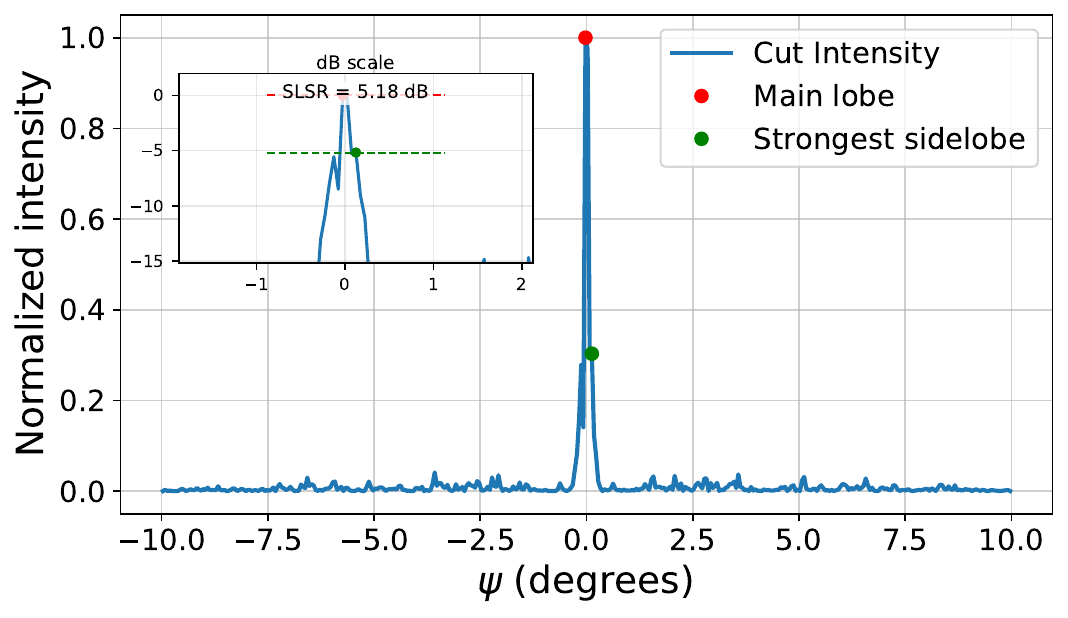}
  }
  \caption{Simulated one-dimensional far-field patterns for the Fibonacci-prime array with 93 antennas in configuration 2: (a) $\theta$-direction (b) $\psi$-direction. Red and green markers indicate the main lobe and the strongest side lobe, respectively. Insets display magnified views of the strongest side lobe region plotted in a logarithmic (dB) scale.}
  \label{fig:6}
\end{figure}

The SLSR is another key performance metric in evaluating an OPA, since a high SLSR is essential for various applications, including LiDAR, and Free Space Optical Communication(FSOC) \cite{N11}. In the proposed design, the combination of non-redundancy and aperiodicity yields a high SLSR. Figures \ref{fig:5a} and \ref{fig:5b} depict the 1-dimensional far-field intensities separately in $\theta$ and $\psi$ directions, respectively, for the arrangement of 93 antennas in Configuration 1. From this, we obtained the SLSRs of 9.64 dB in $\theta$ direction and 13.78 dB in $\psi$ direction. The reason behind the different SLSRs along the two directions lies in the asymmetry of the array arrangement along the x and y axes. A similar analysis for Configuration 2 results in SLSRs of 12.09 dB and 5.18 dB in $\theta$ and $\psi$ directions, respectively. The low SLSR in the $\psi$ direction is attributed to a bulge in the main lobe, which behaves like a side lobe, as shown in Fig.\,\ref{fig:6b}. This deformation, appearing at a higher $\alpha$ value ($\alpha~=~ 0.75$), is linked to the irregularity in the angular spacing of the antennas, leading to a partial interference near the main lobe.

\begin{figure}[htbp]
  \centering

  \subfloat{
    \begin{minipage}[b]{0.3\textwidth}
      \centering
      \includegraphics[width=\linewidth]{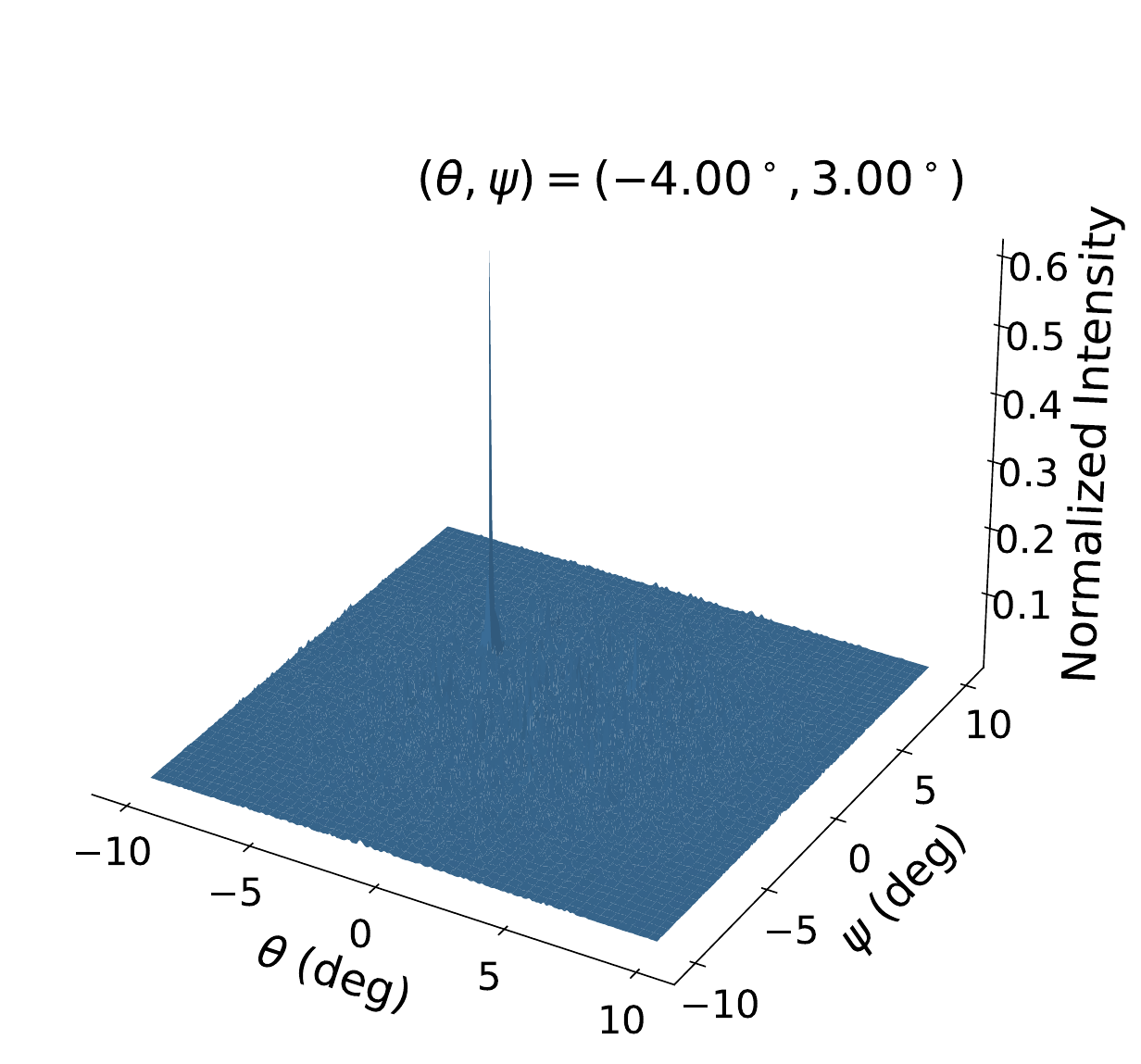}
      \vspace{0.3em}
      \small (a) $\theta_{\text{steer}} = -4.00^\circ $, $\psi_{\text{steer}} = 3.00^\circ $
      \label{fig:7a}
    \end{minipage}
  }
  \hfill
  \subfloat{
    \begin{minipage}[b]{0.3\textwidth}
      \centering
      \includegraphics[width=\linewidth]{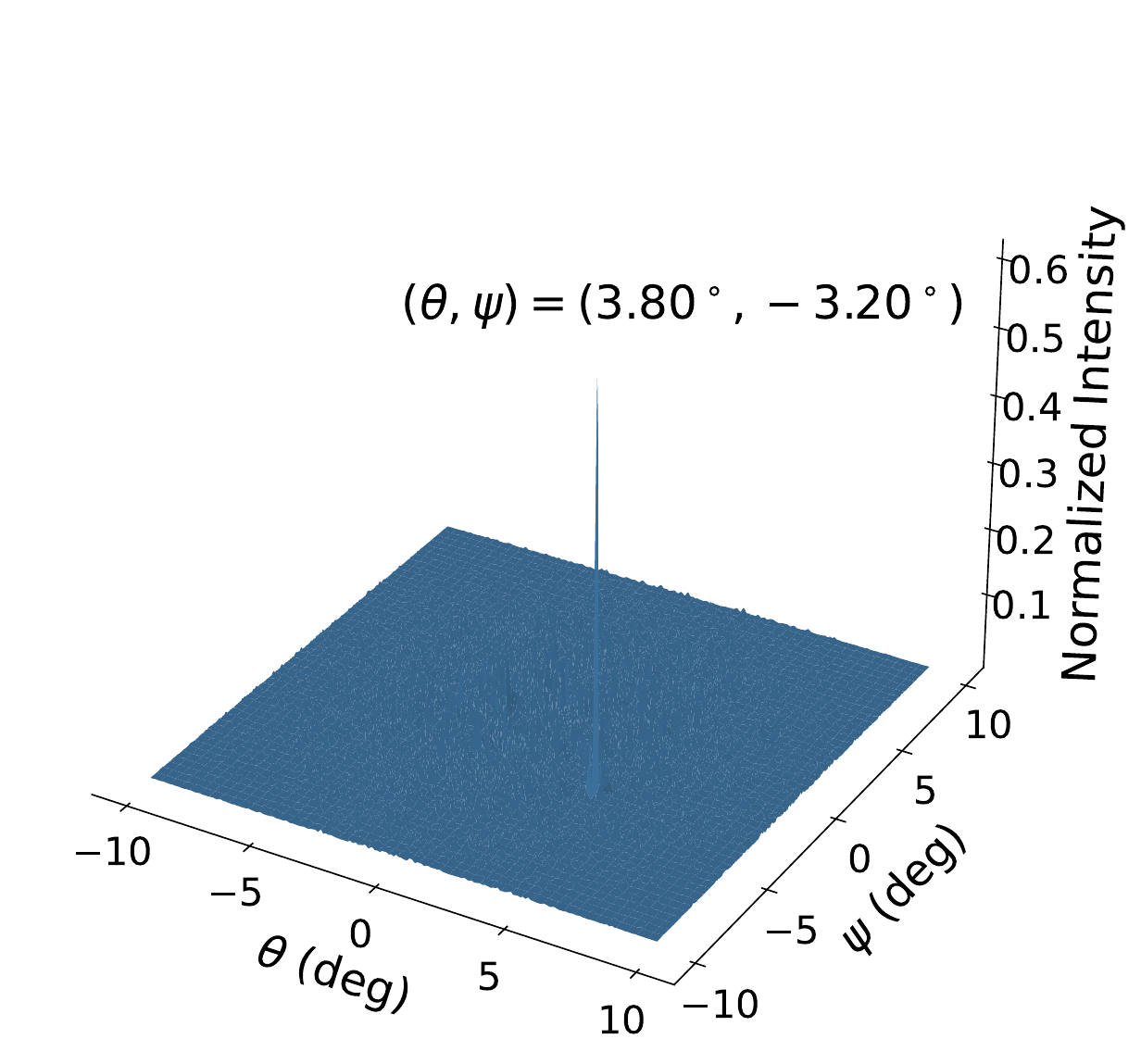}
      \vspace{0.3em}
      \small (b) $\theta_{\text{steer}} = 3.80^\circ$, $\psi_{\text{steer}} = -3.20^\circ $
      \label{fig:7b}
    \end{minipage}
  }
  \hfill
  \subfloat{
    \begin{minipage}[b]{0.3\textwidth}
      \centering
      \includegraphics[width=\linewidth]{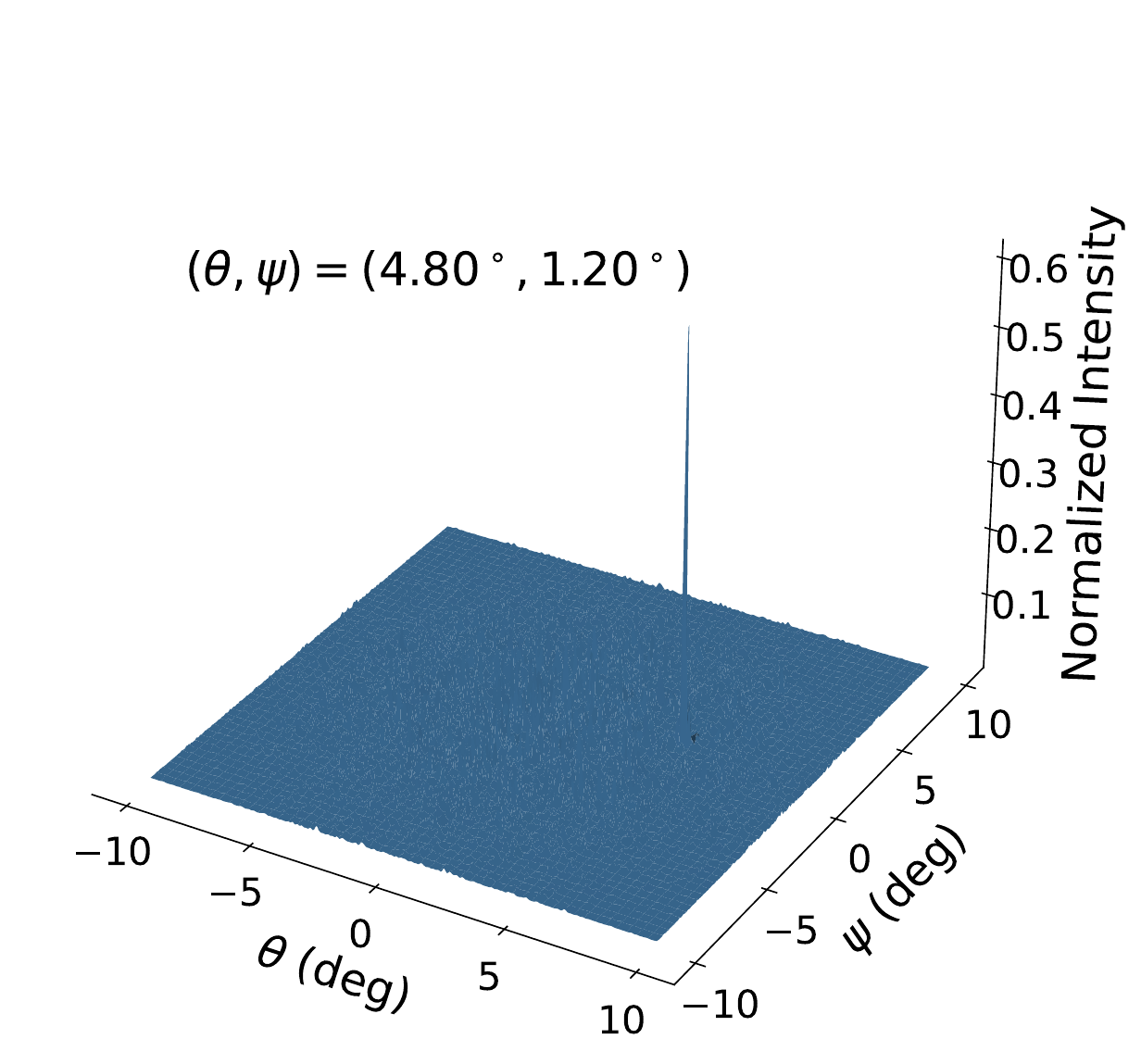}
      \vspace{0.3em}
      \small (c) $\theta_{\text{steer}} = 4.80^\circ $, $\psi_{\text{steer}} = 1.20^\circ$
      \label{fig:7c}
    \end{minipage}
  }

  \vspace{0.4cm}  

  \subfloat{
    \begin{minipage}[b]{0.3\textwidth}
      \centering
      \includegraphics[width=\linewidth]{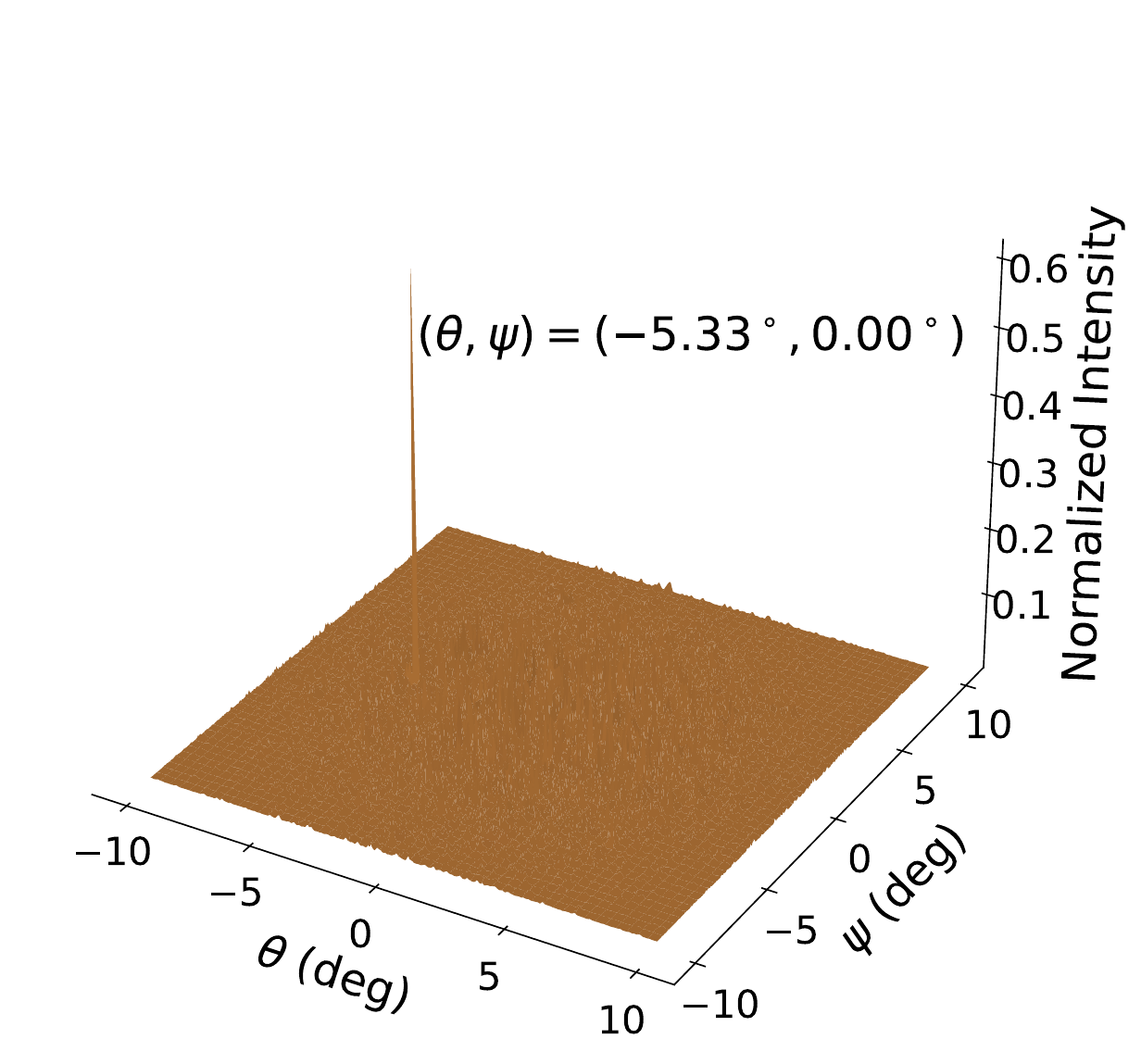}
      \vspace{0.3em}
      \small (d) $\theta_{\text{steer}} = -5.33^\circ $, $\psi_{\text{steer}} = 0.00^\circ$
      \label{fig:7d}
    \end{minipage}
  }
  \hfill
  \subfloat{
    \begin{minipage}[b]{0.3\textwidth}
      \centering
      \includegraphics[width=\linewidth]{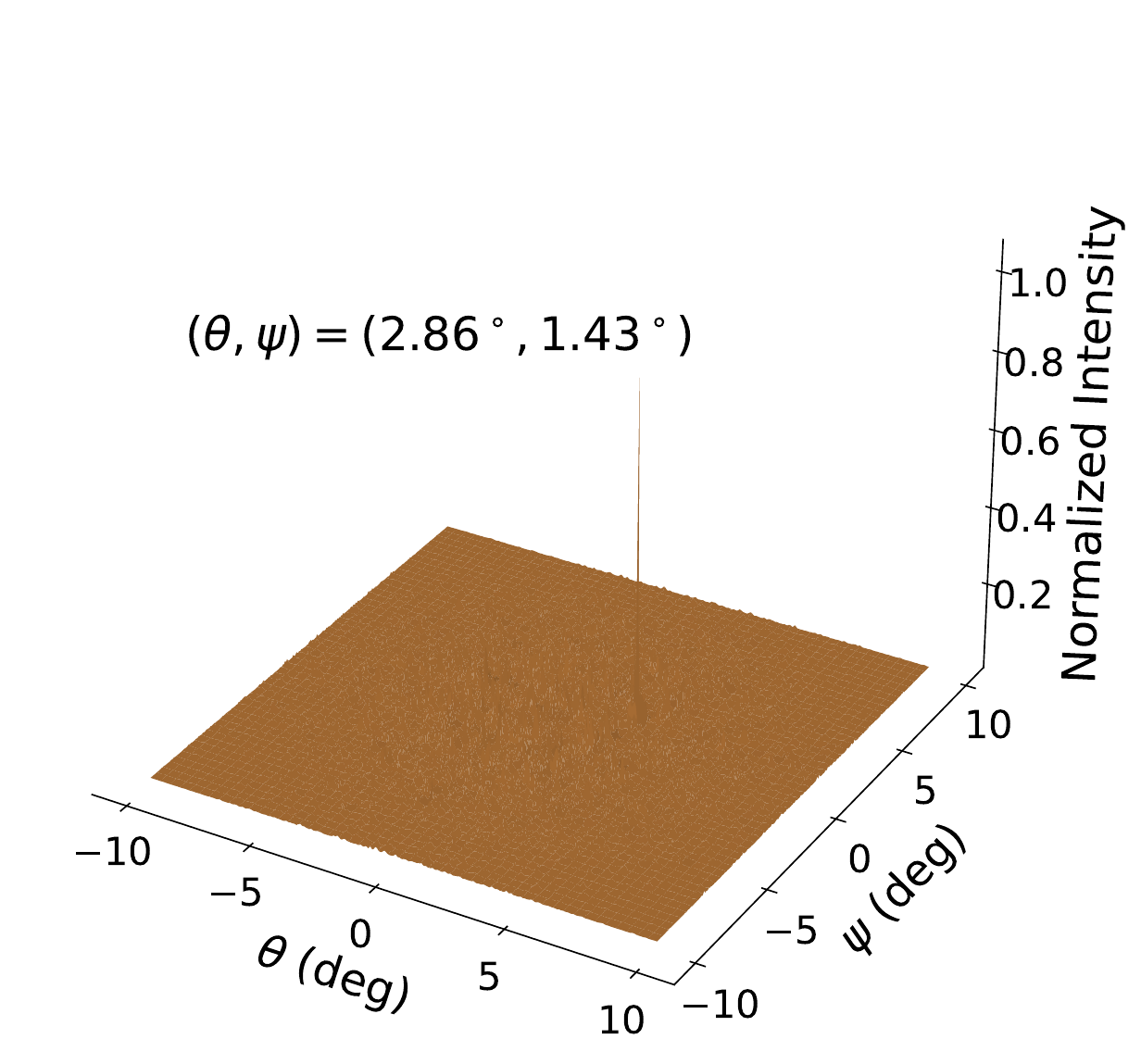}
      \vspace{0.3em}
      \small (e) $\theta_{\text{steer}} = 2.86^\circ $, $\psi_{\text{steer}} = 1.43^\circ $
      \label{fig:7e}
    \end{minipage}
  }
  \hfill
  \subfloat{
  \raisebox{0.35cm}{

    \begin{minipage}[b]{0.3\textwidth}
      \centering
      \includegraphics[width=\linewidth]{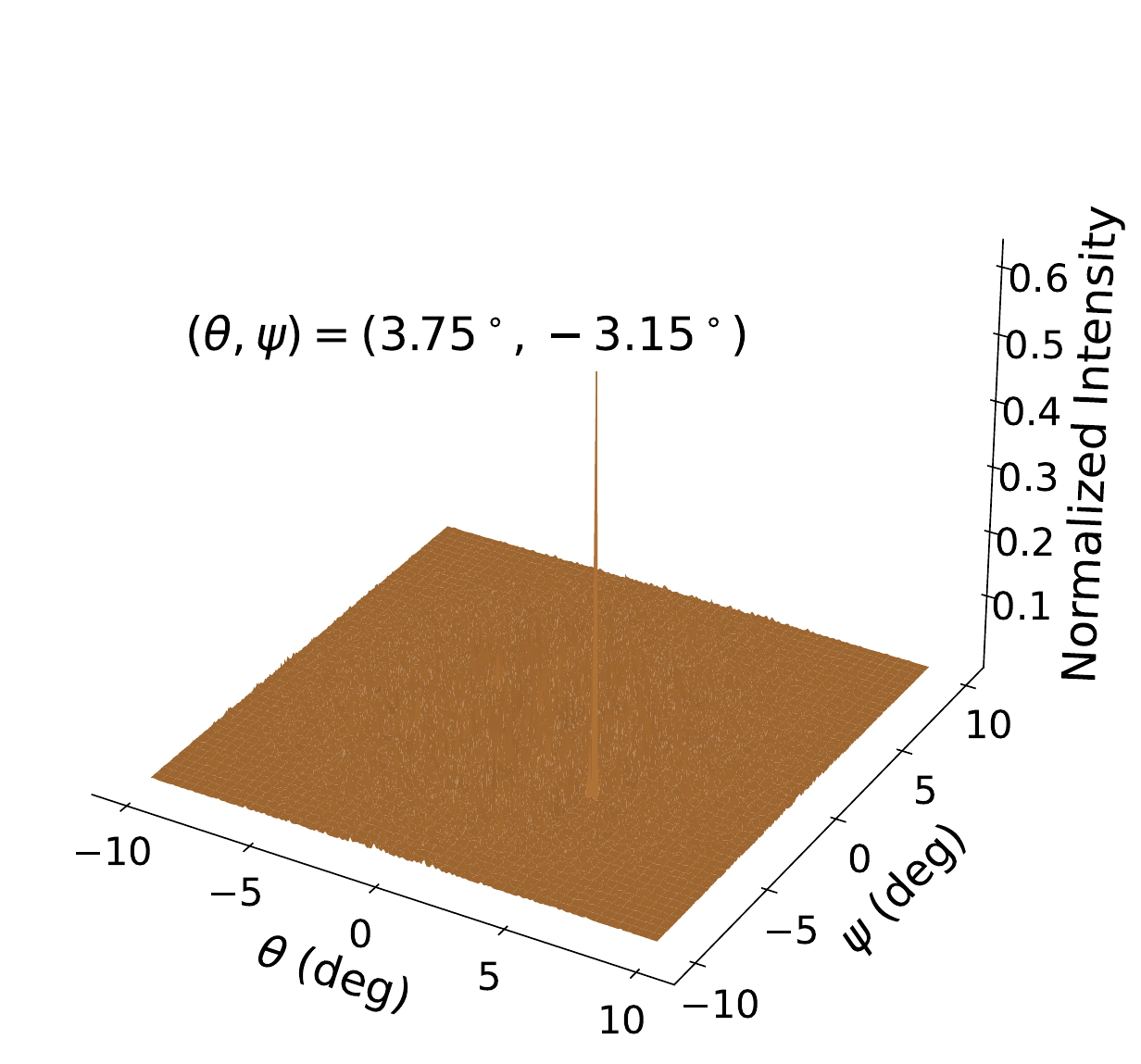}
      \vspace{0.3em}
      \small (f) $\theta_{\text{steer}} = 3.75^\circ $, $\psi_{\text{steer}} = -3.15^\circ $
      \label{fig:7f}
    \end{minipage}
    }
  }

  \caption{Simulated far-field beam steering at different steering angles. Subplots (a-c) correspond to beam steering for the 93-antenna Fibonacci-prime array in configuration 1, while (d-f) correspond to configuration 2.}
  \label{fig:7}
\end{figure}

One of the key challenges for OPAs is achieving a large Field of View (FoV) and a high beam steering, since this requires a sub-wavelength antenna spacing, which leads to cross-coupling between waveguides and large grating lobes \cite{N11, N12}. The aperiodic arrangement of the antennas is one of the strategies developed to suppress the grating lobes and maintain a large steering range \cite{N6, N8, N13, N14}. For a uniform and periodic OPA, the FoV can be calculated directly considering the standard grating lobe limit given by:
\begin{equation}
\text{FoV}_{\text{uniform}} = 2 \sin^{-1}\left( \frac{\lambda}{2d_{\text{min}}} \right)
\label{eq:11}
\end{equation}
where, \( d_{\text{min}} \) is the minimum separation between two antennas. However, for a non-uniform OPA, this will not provide the practical FoV due to various reasons like non-uniform amplitudes/phases, side lobes, non-uniformity of the arrangement, loss of beam quality at extreme angles, etc. Therefore, to account for all these practical factors, which are limiting the FoV, in this work,  we have calculated the FoV by steering the far-field pattern and looking for the angle at which the peak intensity of the main lobe drops to half of the peak intensity of the main lobe in the unsteered condition. This approach visualizes the effective beam coverage to estimate the practical FoV. The resulting FoV for the arrangement of 93 antennas in Configuration 1 is $11.86^\circ \times 11.86^\circ$ in $\theta,\psi$ directions, respectively. On the other hand, the FoV for Configuration 2 turned out to be $11.89^\circ \times 11.89^\circ$. The marginally larger FoV in Configuration 2 results from the fact that a higher $\alpha$ induces a greater angular spread, providing a larger beam steering range.  Fig.\,\ref{fig:7} shows examples of beam steering within the field of view. The top row (Figures \ref{fig:7a}-\ref{fig:7c}) represents the beam steering examples obtained for Configuration 1, while the bottom row (Figures \ref{fig:7d}-\ref{fig:7f}) corresponds to Configuration 2.

The number of resolvable points is defined by the maximum number of distinct spatial points or directions into which the OPA can steer the beam. It is calculated using the equation:
\begin{equation}
    N_{o} ~=~ \frac{\text{FoV}}{\text{FWHM}}
    \label{eq:12}
\end{equation}
Using Equation \ref{eq:12}, the number of resolvable points for 93 antennas in Configuration 1 turns out to be $\approx$ 14,086, and that for Configuration 2 is $\approx$ 56,562. The large number of resolvable points in the case of Configuration 2 originates from the optimization of the control parameter $\alpha$ with the condition to minimize the FWHM and maximize the FoV. This optimization led to an $\alpha ~=~ 0.75$ that provides the maximum number of resolvable points. The optimization of $\alpha$ for Configuration 1 was performed to get a high SLSR, which subsequently caused a reduction in the number of resolvable points. The optimal $\alpha$ values were determined by sweeping $\alpha ~\in~ [0,1]$ in increments of 0.05. Henceforth, depending on the requirement, it is possible to find an optimal Fibonacci-prime arrangement via the optimization of $\alpha$, which will be a tradeoff between the SLSR and the number of resolvable points. 

\section{Experimental Outlook}
\label{sec:4}

\begin{figure}[htbp]
  \centering

  \subfloat{
    \begin{minipage}[b]{0.48\textwidth}
      \centering
      \includegraphics[width=\linewidth]{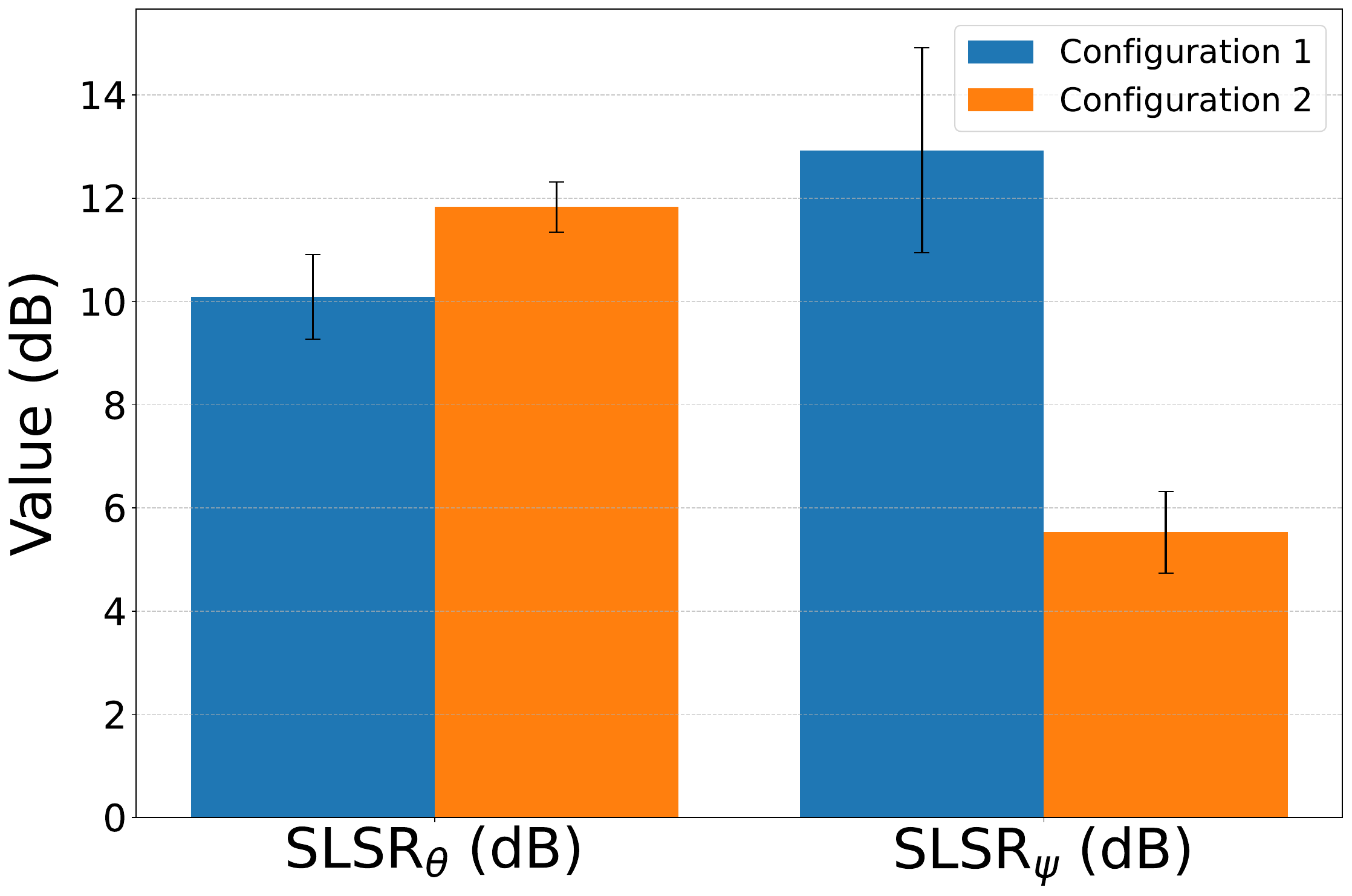}
      \vspace{0.3em}
      \small (a)
      \label{fig:8a}
    \end{minipage}
  }
  \hfill
  \subfloat{
    \begin{minipage}[b]{0.48\textwidth}
      \centering
      \includegraphics[width=\linewidth]{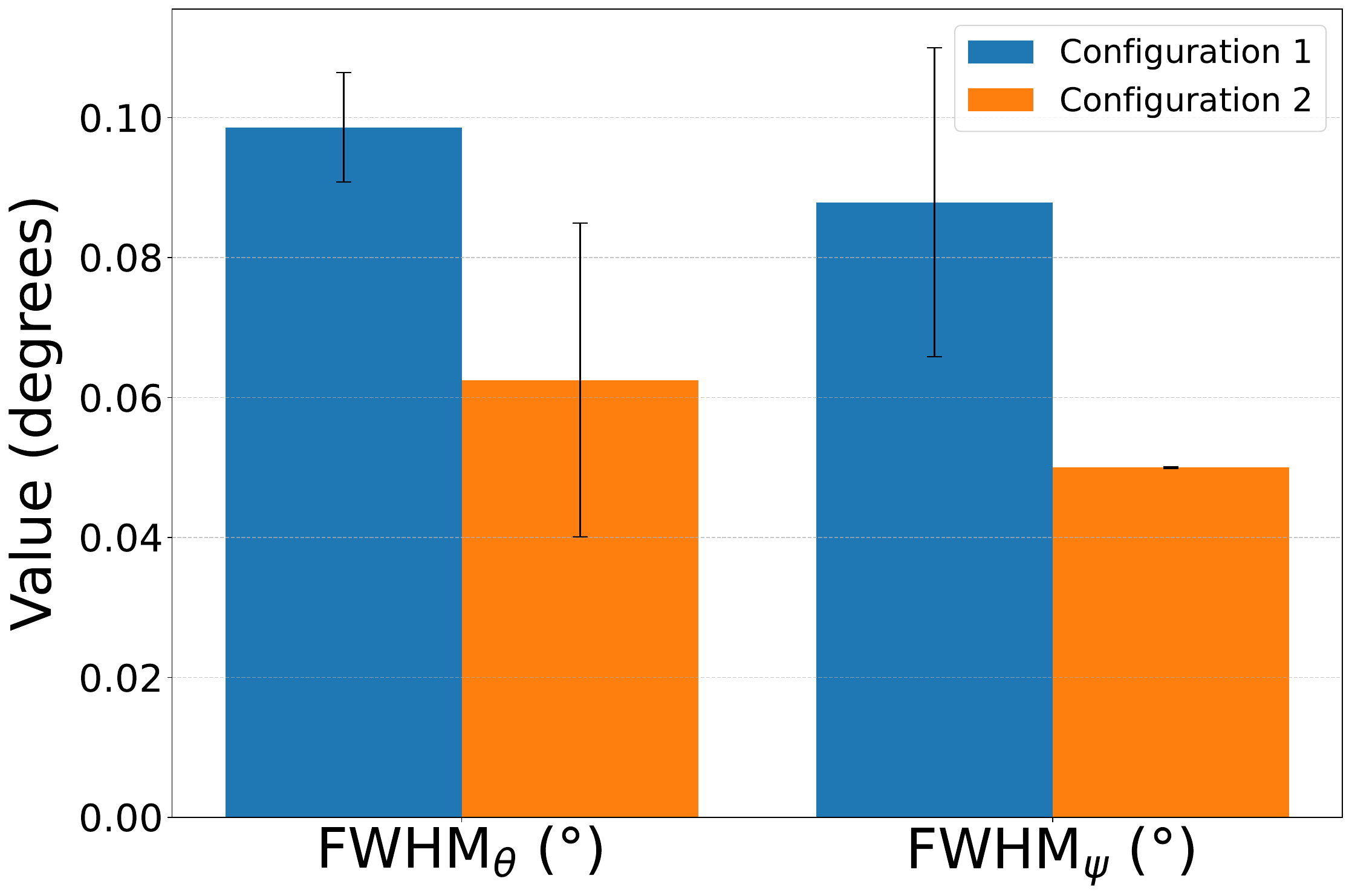}
      \vspace{0.3em}
      \small (b) 
      \label{fig:8b}
    \end{minipage}
  }

  \vspace{0.6cm} 

  \subfloat{
    \begin{minipage}[b]{0.48\textwidth}
      \centering
      \includegraphics[width=\linewidth]{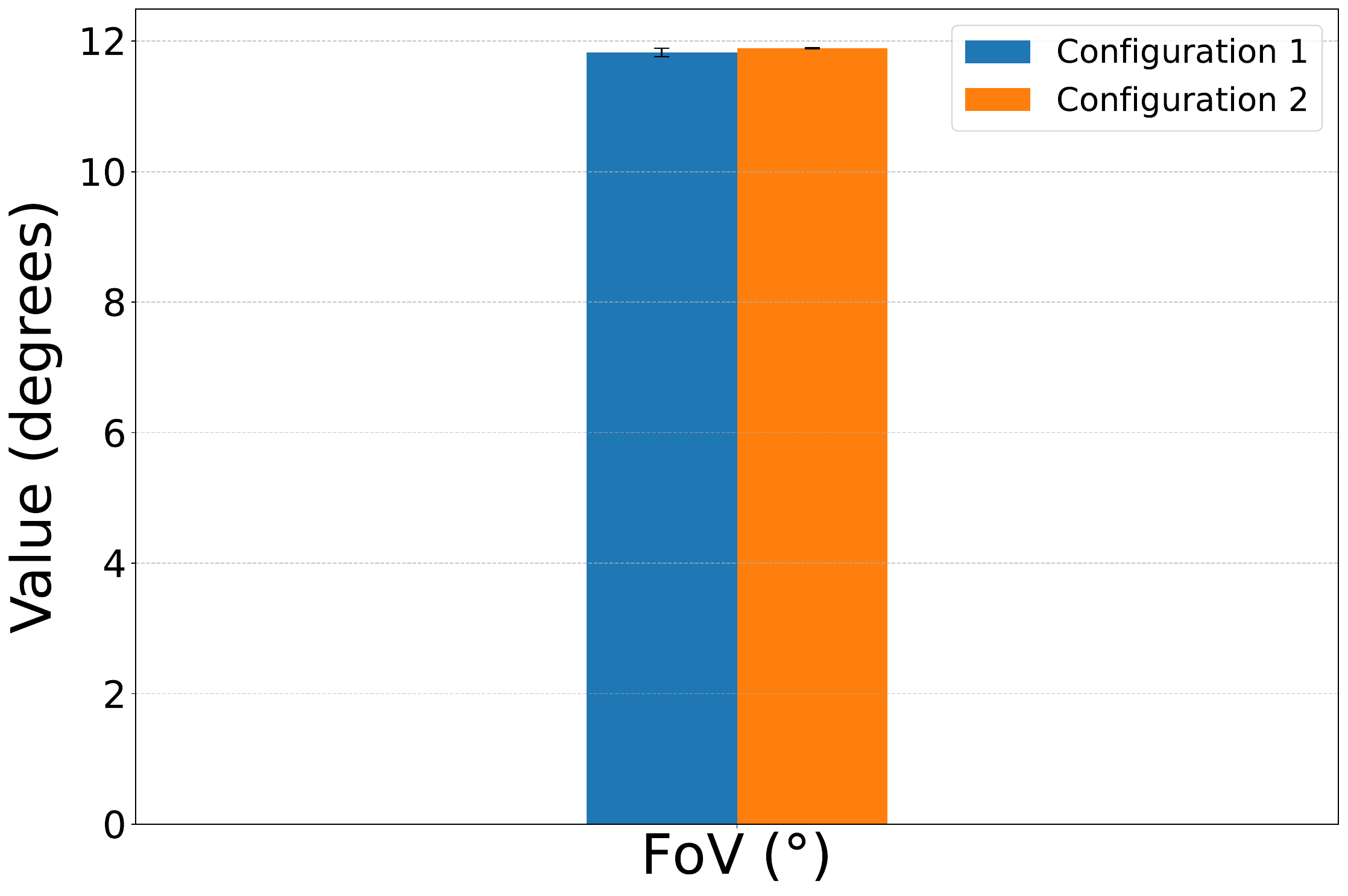}
      \vspace{0.3em}
      \small (c) 
      \label{fig:8c}
    \end{minipage}
  }

  \caption{Comparison of (a) side-lobe suppression ratio (SLSR), (b) full width at half-maximum (FWHM), and (c) field of view (FoV)  for the two Fibonacci-prime array configurations with 93 antennas. Error bars indicate variations obtained from multiple simulation runs.}
  \label{fig:8}
\end{figure}

As an outlook for experimental validation, we tested the tolerance of this design to the variations in antenna positioning, which, for instance, may originate from the fabrication process for the on-chip realization of OPAs. We have introduced normally distributed offsets with a standard deviation $\sigma ~=~0.04$ to the device parameters $a_p$ and $\theta_p$, and consequently analyzed how the far-field performance metrics: SLSR, FWHM, and FoV, vary with respect to these offsets. Table \ref{tab:1} details a statistical summary of the far-field metrics calculated over 400 iterations. We find that the deviations of antenna positions have a varied impact on different far-field parameters, and this also depends on the antenna configurations. Notably, the SLSR is maximally impacted by the antenna position disturbances compared to the FWHM and the FoV. Moreover,
the standard deviations of SLSR are higher along the $\psi$-direction compared to the $\theta$-direction. A similar trend is also observed for the FWHM at the far-field. This is linked to the fact that the antenna distributions are inherently uneven in the proposed Fibonacci-prime geometry, leading to a larger impact of positional disturbance towards the $\psi$-axis compared to that along the $\theta$-axis. We find that at 4$\%$ offset, for configuration 1, the standard deviation in SLSR along the $\psi$- direction, SLSR\textsubscript{$\psi$}, approaches nearly 2 dB. On the other hand, the SLSR is relatively more robust against perturbations of antenna positioning for configuration 2, although for this configuration, the SLSR\textsubscript{$\psi$} is significantly lower compared to configuration 1. Fig.\,\ref{fig:8} provides a visualization of these statistics, which confirms that Configuration 2 achieves maximum beam confinement with lower variability, and Configuration 1 provides the highest SLSR in the $\psi$ direction, though highly sensitive to any antenna positional deviations. The variations in antenna positioning least perturb the FoV for both configurations. These results indicate that moderate antenna positioning errors due to fabrication imperfections ($\leq4\%$) can be tolerated without significant performance degradation. For the considered minimum antenna spacing of 15 $\mu$m, a 4\% positional deviation corresponds to $\approx$0.6 $\mu$m offset, which is well within the reach of state-of-the-art fabrication process tolerances.

\begin{table}[h!]
\centering
\caption{Performance comparison along with tolerance to fabrication errors of configuration-1 and configuration-2 for key far-field metrics-- side lobe suppression ratio (SLSR), full width at half-maximum (FWHM), and field of view (FoV). Values are expressed as mean $\pm$ standard deviation.}
\begin{tabular}{lcc}
\toprule
\textbf{Metric} & \textbf{Configuration-1} & \textbf{Configuration-2} \\
\midrule
SLSR$_{\theta}$ (dB)  & $10.09 \,\pm\, 0.82$   & $11.83 \,\pm\, 0.49$ \\
SLSR$_{\psi}$ (dB)    & $12.93 \,\pm\, 1.99$   & $5.53 \,\pm\, 0.79$  \\
FWHM$_{\theta}$ (°)   & $0.0986 \,\pm\, 0.0078$ & $0.0625 \,\pm\, 0.0224$ \\
FWHM$_{\psi}$ (°)     & $0.0879 \,\pm\, 0.0221$ & $0.0500 \,\pm\, 0.0001$ \\
FoV (°)               & $11.8265 \,\pm\, 0.0658$ & $11.8909 \,\pm\, 0.006$ \\
\bottomrule
\end{tabular}
\label{tab:1}
\end{table}

\begin{figure}[htbp]
  \centering
  \subfloat[\label{fig:9a}]{
    \includegraphics[width=0.48\textwidth]{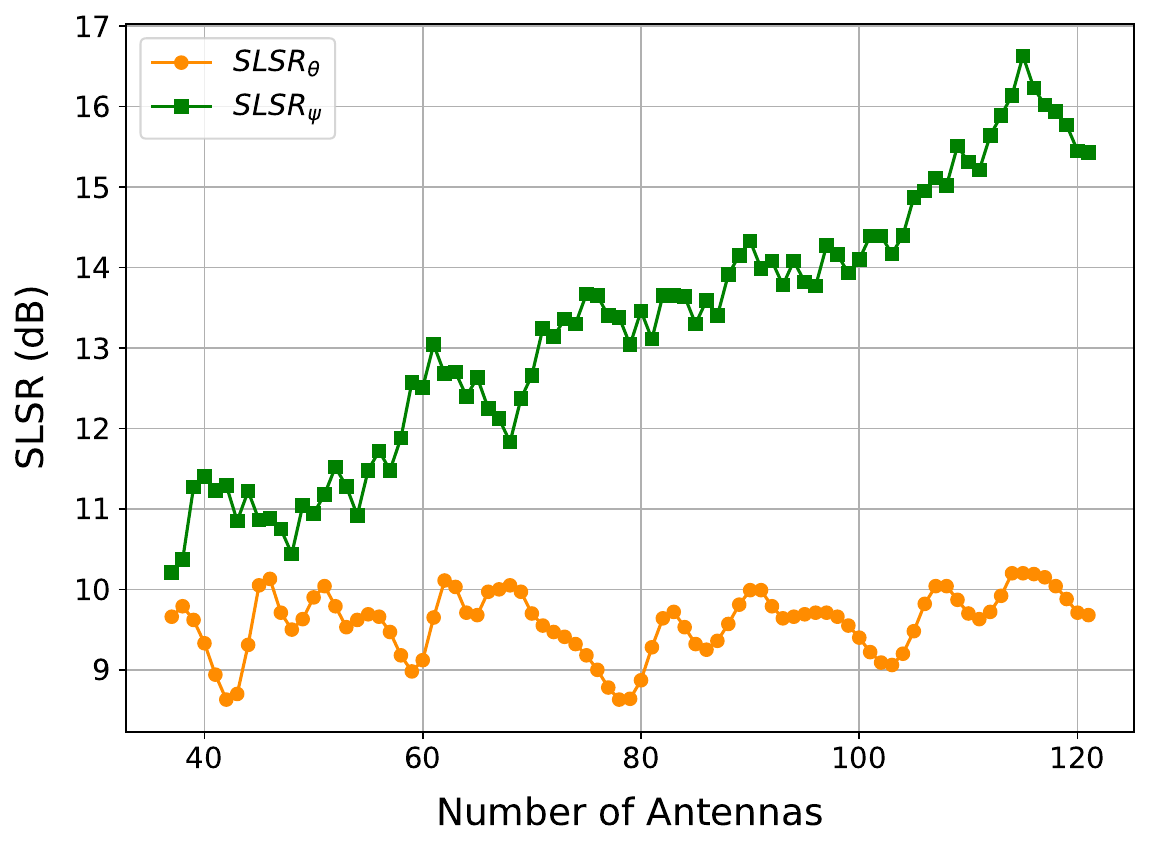}
  }
  \hfill
  \subfloat[\label{fig:9b}]{
    \includegraphics[width=0.48\textwidth]{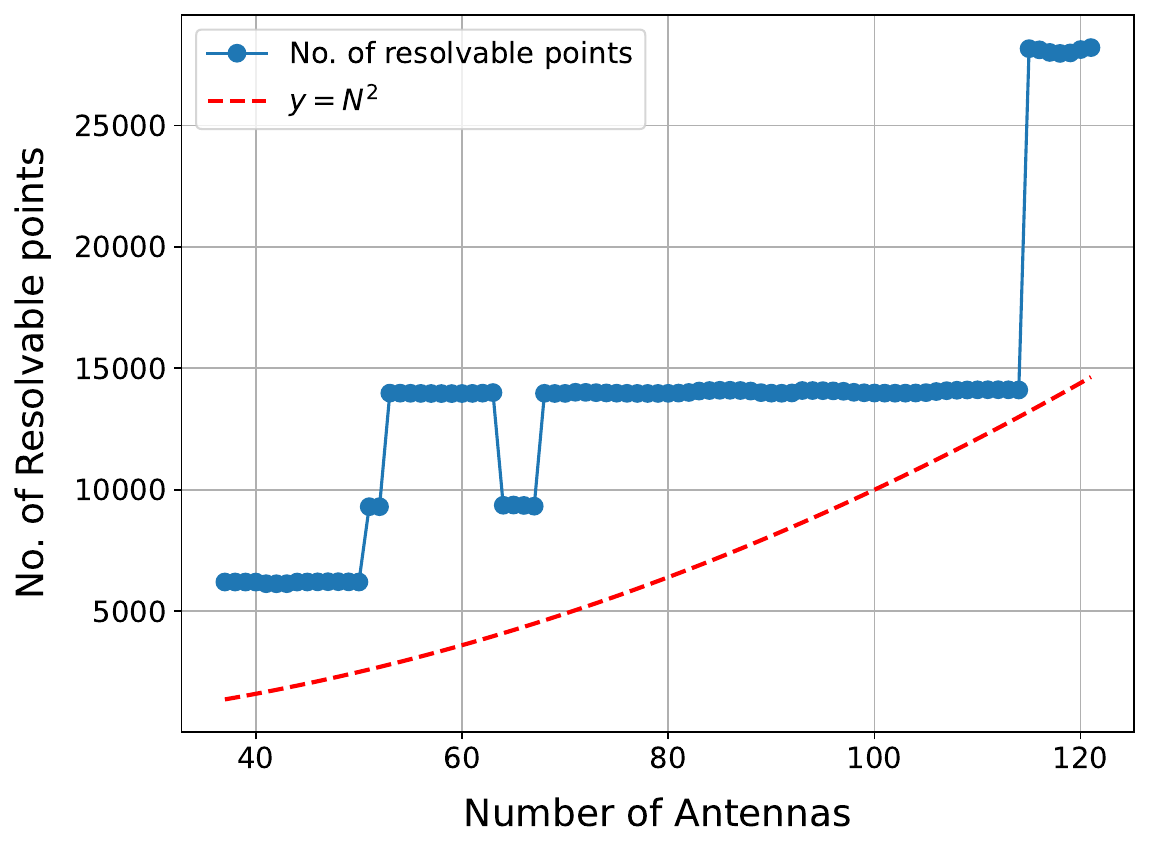}
  }
  \caption{Trend analysis for configuration 1 (a) Variation of the side lobe suppression ratio (SLSR) with the number of antennas, where orange and green dots correspond to $\theta$- and $\psi$- directions, respectively (b) Trend of the number of resolvable points (blue dots) compared with the theoretical $y ~=~ N^2$ curve(red dashed line). Minor discontinuities arise from the discrete addition of antennas in the Fibonacci-prime geometry.}
  \label{fig:9}
\end{figure}

 To analyze the scalability of this OPA design, we evaluated the far-field performance metrics as a function of the number of antennas. Fig.\,\ref{fig:9} represents the trend analysis of Configuration 1 with an increasing number of antennas, where Fig.\,\ref{fig:9a} represents the variation of the SLSR along both $\psi-$ and $\theta-$directions, and Fig.\,\ref{fig:9b} demonstrates the variation of the number of resolvable points. Fig.\,\ref{fig:9b} shows an approximate saturation in the number of resolvable points with increasing number of antennas from around 55 to 110. This early saturation can be attributed to the lower $\alpha$ value (0.45), which produces a denser but less steerable aperture, limiting the incremental gain in FoV with additional antennas. Fig.\,\ref{fig:9a}, however, shows a steady increase in $SLSR_\psi$, whereas SLSR along $\theta-$direction maintains a constant value. This unique behaviour offers design flexibility: since the number of resolvable points remains nearly constant, applications demanding stronger SLSR can employ more antennas, while those prioritizing compactness can use fewer elements without compromising on angular resolution. 
 
\begin{figure}[htbp]
  \centering
  \subfloat[\label{fig:10a}]{
    \includegraphics[width=0.48\textwidth]{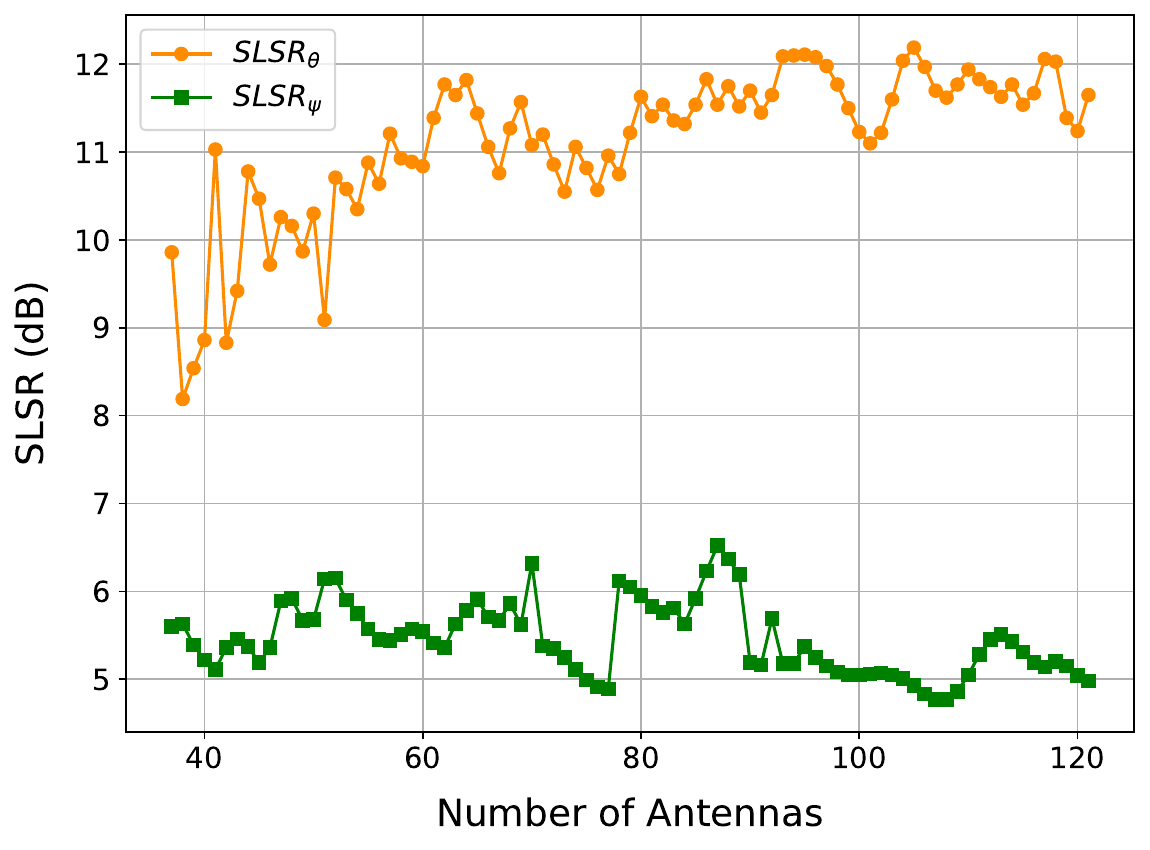}
  }
  \hfill
  \subfloat[\label{fig:10b}]{
    \includegraphics[width=0.48\textwidth]{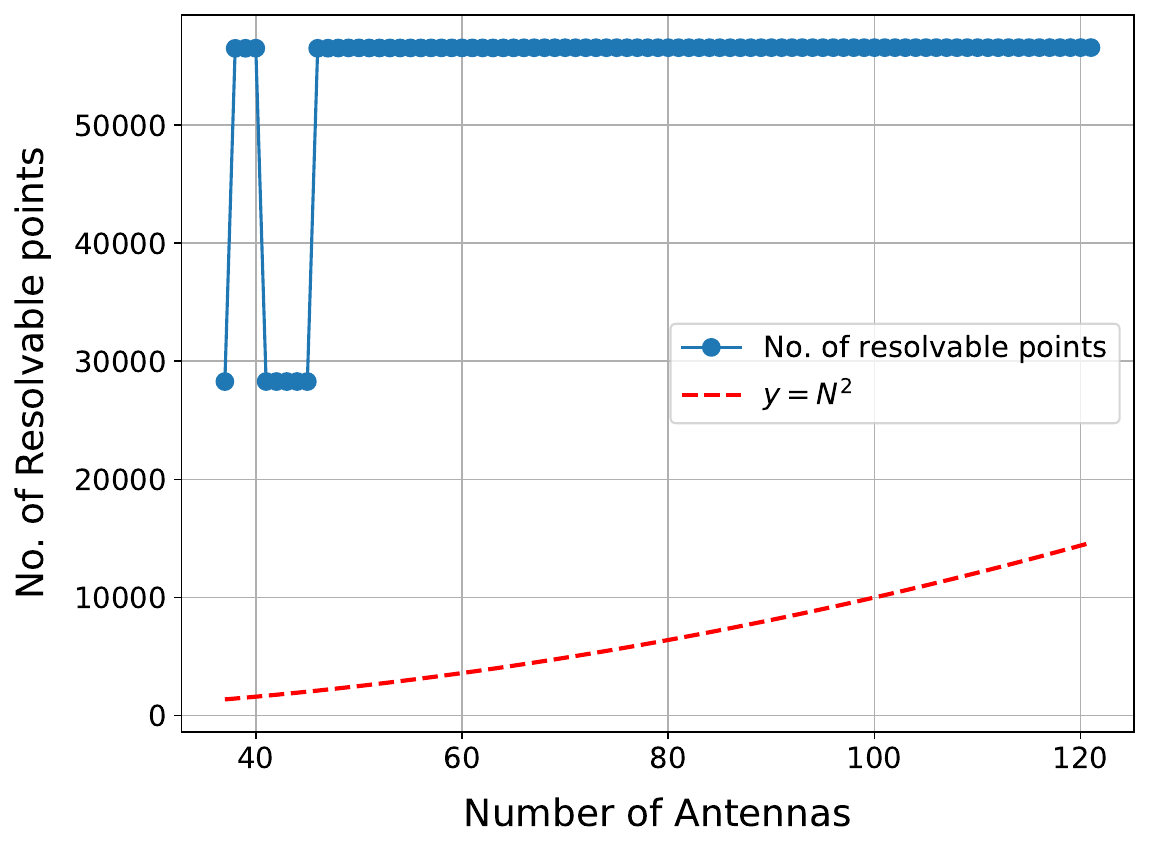}
  }
  \caption{Trend analysis for configuration 2 (a) Variation of the side lobe suppression ratio (SLSR) with the number of antennas, where orange and green dots correspond to $\theta$- and $\psi$- directions, respectively (b) Dependence of the number of resolvable points (blue dots) compared with the theoretical $y ~=~ N^2$ curve (red dashed line). Step like changes reflect discrete geometric transitions as additional antennas are introduced.}
  \label{fig:10}
\end{figure}

\begin{figure}[htbp]
  \centering
  \subfloat[\label{fig:11a}]{
    \includegraphics[width=0.48\textwidth]{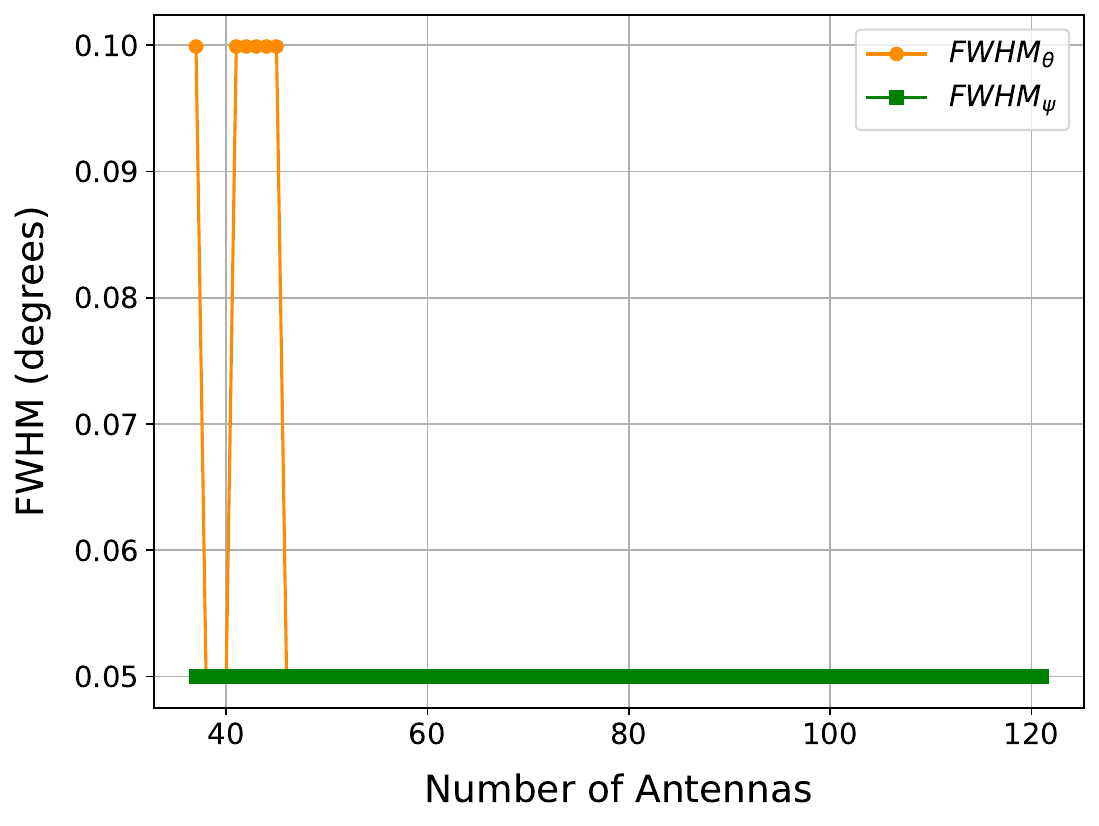}
  }
  \hfill
  \subfloat[\label{fig:11b}]{
    \includegraphics[width=0.48\textwidth]{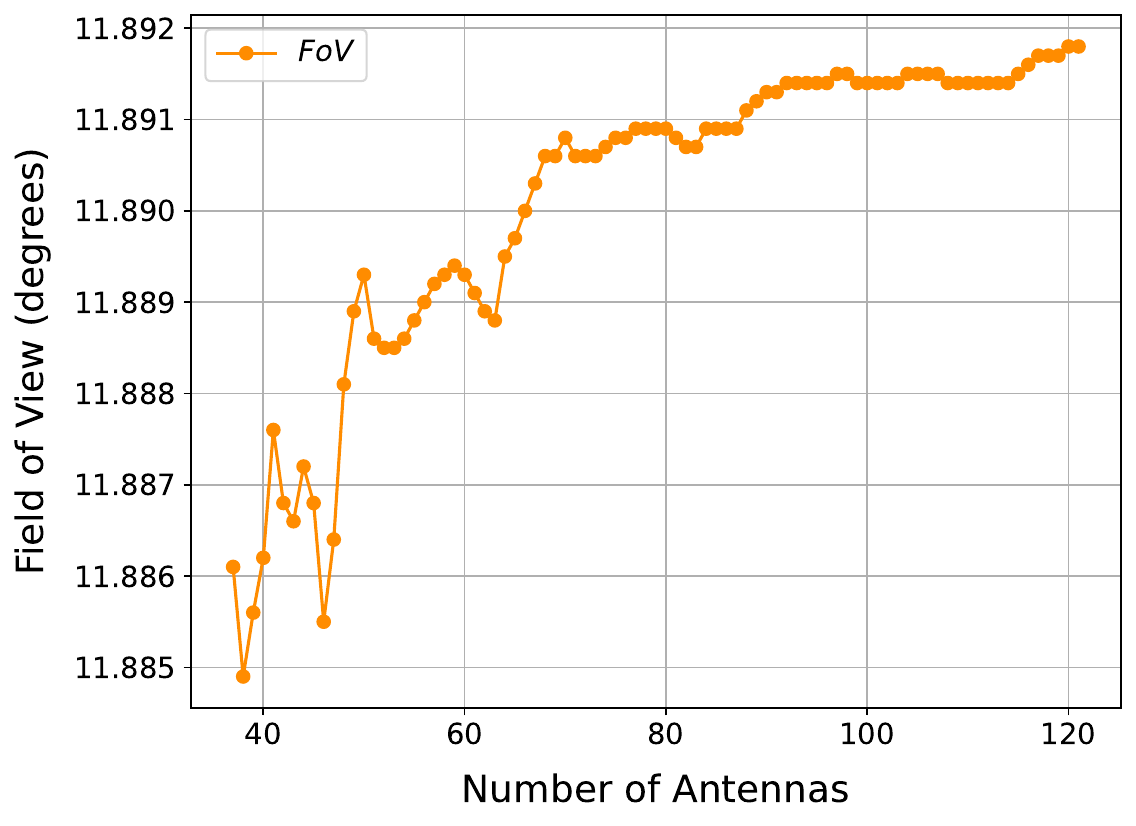}
  }
  \caption{Variation of (a) full width at half-maximum (FWHM), Orange and green dots represent values along $\theta$- and $\psi$-directions, respectively (b) field of view (FoV) with increasing number of antennas in Configuration 2}
  \label{fig:11}
\end{figure}

 This scaling trend analysis with the number of antennas for Configuration 2 is shown in Fig.\,\ref{fig:10}. Clearly, from Fig.\,\ref{fig:10a}, the SLSRs in both $\theta$ and $\psi$ directions exhibit a saturation behaviour. Interestingly, as depicted in Fig.\,\ref{fig:10b}, almost a constant number of resolvable points is achieved throughout the range of scaling. This saturation can be understood from Fig.\,\ref{fig:11}, which shows the variation of FWHM and FoV with increasing the number of antennas. Fig.\,\ref{fig:11a} displays a constant FWHM in both $\theta$ and $\psi$ directions. Although Fig.\,\ref{fig:11b} shows an increasing trend in FoV, the order of this increase is very low($\approx0.007^\circ$). Since the number of resolvable points depends on the ratio of the FoV to the FWHM, saturation in these values led to almost constant resolvable points.\\
 The key observation, however, is on the order of the values. A large number of resolvable points of around $\approx$ 55,000 clearly shows that this Configuration is highly optimized at minimizing the FWHM and maximizing FoV. Figures \ref{fig:9b} and \ref{fig:10b} show a step-like change in the number of resolvable points ($N_o$ = FoV/FWHM) as the antenna count is increased. This is due to the requirement of the addition of extra arcs into the proposed Fibonacci-prime arrangement to accommodate a higher number of antennas, which allows for only a discrete change in the number of antennas. This creates small variations in the FoV and FWHM, leading to abrupt changes in the FoV/FWHM ratio.  
 
To give a fair comparison of the proposed Fibonacci-prime OPA design with the non-redundant Costas Array design by T.Fukui \textit{et al}. \cite{N7}, we collated the far-field metrics of our Fibonacci spiral-based arrangement of 121 antennas in Configuration 1 and 2 with those of the Costas array with 127 antennas in Table \ref{tab:2}. The FWHM of the Costas array is slightly better than both configurations of our Fibonacci-prime arrangement. Nonetheless, the present Fibonacci-prime design clearly outperforms the Costas arrangement in the FoV aspect, and hence provides a much higher number of resolvable points. This improvement stems from the Fibonacci spiral's quasi-continuous radial expansion, which preserves wide angular coverage compared to the strictly discrete Costas pattern. Additionally, the novelty of our proposed design lies in the ability to provide tunable trade-offs via the $\alpha$ parameter and a simpler algorithm for generating the layout, unlike the fixed Costas array layout. 

\begin{table}[h!]
\centering
\caption{Comparison of performance metrics for the Fibonacci-prime array configurations and Costas Array \cite{N7}.}
\begin{tabular}{lccc}
\toprule
\textbf{Metric} & \textbf{Configuration-1} & \textbf{Configuration-2} & \textbf{Costas Array \cite{N7}} \\
\midrule
No. of Antennas & 121 & 121 & 127 \\
FWHM (\(^\circ\)) & $0.999 \times 0.050$ & $0.050 \times 0.050$ & $0.0428 \times 0.0428$ \\
FoV (\(^\circ\)) & $11.87 \times 11.87$ & $11.89 \times 11.89$ & $5.92 \times 5.92$ \\
SLSR$_{\theta}$ (dB) & 9.68 & 11.65 & $\sim$ 12 \\
SLSR$_{\psi}$ (dB) & 15.43 & 4.98 & $\sim$ 12 \\
No. of Resolvable Points & $\sim 28,200$ & $\sim 56,500$ & $\sim 19,000$ \\
\bottomrule
\end{tabular}
\label{tab:2}
\end{table}

\section{Conclusion}
In this work, we have proposed a novel design of an OPA based on a Fibonacci spiral. We have analyzed its performance by evaluating crucial parameters like FWHM, FoV, and SLSR at the far-field. We introduced an antenna-positioning control parameter $\alpha$, which serves as a design knob to tailor OPAs for specific applications through the optimization of a certain far-field parameter. This work, to the best of our knowledge, is the first of its kind to combine prime number parameterization with the control parameter $\alpha$ to achieve high SLSR and a large number of resolvable points by exploiting the versatility of the Fibonacci spiral. We have inspected the scalability of this design with an increasing number of antennas and studied its far-field performances. The proposed Fibonacci Spiral-based OPA design with a minimum antenna separation of $15\,\mu\textrm{m}$ is well within the reach of state-of-the-art fabrication processes for standard integrated photonics platforms, including silicon-on-insulator(SOI) \cite{N15}, silicon nitride \cite{N20}, lithium niobate \cite{N21}, indium gallium aluminum arsenide \cite{N22}, and indium phosphide \cite{N23}. The control parameter $\alpha$, which corresponds to an angular positioning offset, can also be physically realized during antenna placement through standard fabrication methods. We also examined the tolerances of the proposed OPA design to the random variations in the antenna positions, and verified that the present design is quite robust to the expected fabrication errors. In conclusion, the proposed nature-inspired array layout offers a promising pathway toward realizing Optical Phased Arrays for diverse industrial applications like LiDAR, FSOC, image projection, to name a few. 

\section*{Funding}
This research was funded by the Ministry of Electronics and Information Technology (MeitY), under the Government of India, through the "National Centre for Quantum Accelerator Chip Using Lithium Niobate (NCQAC)".
\section*{Disclosures}
The authors declare no conflicts of interest.

\section*{Data Availability}
 Data underlying the results presented in this paper are not publicly available at this time but maybe obtained from the authors upon reasonable request.


\bibliographystyle{unsrt}  
\bibliography{references}

\end{document}